\documentclass[fleqn,usenatbib]{mnras}

\usepackage{newtxtext,newtxmath}

\usepackage[T1]{fontenc}

\DeclareRobustCommand{\VAN}[3]{#2}
\let\VANthebibliography\thebibliography
\def\thebibliography{\DeclareRobustCommand{\VAN}[3]{##3}\VANthebibliography}

\usepackage{graphicx}	
\usepackage{amsmath}	

\newcommand{\Ehs}{E_{\mathrm{hs}}}
\newcommand{\Ebin}{E_{\mathrm{b}}}
\newcommand{\rhz}{r_{\mathrm{h},0}}
\newcommand{\rc}{r_{\mathrm{c}}}
\newcommand{\rh}{r_{\mathrm{h}}}
\newcommand{\rhbh}{r_{\mathrm{h,BH}}}
\newcommand{\nhb}{\overline{N}_{\mathrm{hb}}}

\newcommand{\rhn}{R_{\mathrm{hn}}}
\newcommand{\fbh}{f_{\mathrm{BH}}}
\newcommand{\Mbh}{M_{\mathrm{BH}}}
\newcommand{\trh}{t_{\mathrm{rh}}}
\newcommand{\trha}{t_{\mathrm{rh1}}}
\newcommand{\fbin}{f_{\mathrm{bin}}}

\newcommand{\mbv}{m_{\mathrm{obs}}}
\newcommand{\mb}{m_{\mathrm{tot}}}

\newcommand{\dvr}{|\Delta v_\mathrm{LOS}|}
\newcommand{\cnm}{\overline{N_{\mathrm{f}}}(m)}
\newcommand{\cnmb}{\overline{N_{\mathrm{a}}}(m)}
\newcommand{\vlos}{v_{\mathrm{LOS}}}
\newcommand{\slos}{\sigma_{\mathrm{LOS}}}
\newcommand{\sloss}{\sigma_{\mathrm{LOS,S}}}
\newcommand{\slossrc}{\sigma_{\mathrm{LOS,S,hn}}}
\newcommand{\slosb}{\sigma_{\mathrm{LOS,SB}}}
\newcommand{\sloscb}{\sigma_{\mathrm{LOS,SCB}}}
\newcommand{\slosvirial}{\sigma_{\mathrm{1D}}}
\newcommand{\Tmat}{T_{\mathrm{mat}}}

\usepackage[normalem]{ulem} 

\title[Binary populations of Pal 5]{The influence of black holes on the binary population of the globular cluster Palomar~5}

\author[Long Wang et al.]{
Long Wang,$^{1,2}$\thanks{E-mail: wanglong8@sysu.edu.cn (SYSU)}
Mark Gieles,$^{3,4}$
Holger Baumgardt,$^{5}$
Chengyuan Li,$^{1,2}$
Xiaoying Pang,$^{6,7}$
and Baitian Tang$^{1,2}$
\\
$^{1}$ School of Physics and Astronomy, Sun Yat-sen University, Daxue Road, Zhuhai, 519082, China\\
$^{2}$ CSST Science Center for the Guangdong-Hong Kong-Macau Greater Bay Area, Zhuhai, 519082, China\\
$^{3}$ Institut de Ci\`{e}ncies del Cosmos (ICCUB), Universitat de Barcelona (IEEC-UB), Mart\'{i} i Franqu\`{e}s 1, E08028 Barcelona, Spain\\
$^{4}$ ICREA, Pg. Llu\'{i}s Companys 23, E08010 Barcelona, Spain\\
$^{5}$ School of Mathematics and Physics, The University of Queensland, St. Lucia, QLD 4072, Australia\\
$^{6}$ Department of Physics, Xi’an Jiaotong-Liverpool University, 111 Ren’ai Road, Suzhou Dushu Lake Science and Education Innovation District,\\~~~Suzhou Industrial Park, Suzhou 215123, P.R. China\\
$^7$ Shanghai Key Laboratory for Astrophysics, Shanghai Normal University, 100 Guilin Road, Shanghai 200234, P. R. China\\
}

\date{Accepted XXX. Received YYY; in original form ZZZ}

\pubyear{2022}

\begin{document}
\label{firstpage}
\pagerange{\pageref{firstpage}--\pageref{lastpage}}
\maketitle

\begin{abstract}

The discovery of stellar-mass black holes (BHs) in globular clusters (GCs) raises  the possibility of long-term retention of BHs within GCs. 
These BHs influence various astrophysical processes, including merger-driven gravitational waves and the formation of X-ray binaries.
They also impact cluster dynamics by heating and creating low-density cores.
Previous $N$-body models suggested that Palomar~5, a low-density GC with long tidal tails, may contain more than 100 BHs. 
To test this scenario, we conduct {\it N}-body simulations of Palomar~5 with primordial binaries to explore the influence of BHs on binary populations and the stellar mass function. 
Our results show that primordial binaries have minimal effect on the long-term evolution. 
In dense clusters with BHs, the fraction of wide binaries with periods >$10^5$ days decreases, and the disruption rate is independent of the initial period distribution. 
Multi-epoch spectroscopic observations of line-of-sight velocity changes can detect most bright binaries with periods below $10^4$~days, significantly improving velocity dispersion measurements.
Four BH-MS binaries in the model with BHs suggests their possible detection through the same observation method. 
Including primordial binaries leads to a flatter inferred mass function because  of spatially unresolved binaries, leading to a better match of the observations than models without binaries,  particularly in Palomar~5's inner region.
Future observations should focus on the cluster velocity dispersion and binaries with periods of $10^4-10^5$ days in Palomar 5's inner and tail regions to constrain BH existence.

\end{abstract}




\section{Introduction}

Following several detections of stellar-mass black hole (BH) candidates through X-ray and radio observations \citep{Strader2012,Chomiuk2013,Miller-Jones2015,Bahramian2017} and via radial velocity measurements \citep{Giesers2018,Giesers2019} in globular clusters (GCs), the long-term dynamical impact of BHs in GCs has been extensively studied \citep[e.g.][]{Breen2013,Morscher2013,Morscher2015,Sippel2013,Heggie2014,Wang2016,Sollima2016,Peuten2016,Rodriguez2016,Askar2018,Weatherford2020,Wang2020,Weatherford2021,Wang2021,Gieles2023}.
Investigating the BH population is also crucial for constraining the massive end of the initial mass function (IMF) \citep[e.g.][]{Shanahan2015,Chatterjee2017,Henault-Brunet2020,Baumgardt2023,Dickson2023}.
\cite{Breen2013} demonstrated that the presence of BH subsystems significantly impacts the evolution of star clusters, with BHs forming binary BHs (BBHs) and controlling the central energy flow. 
\cite{Wang2020}  further showed that a large fraction of BHs would accelerate the relaxation process and leads to faster tidal disruption of GCs. 
In the case of a top-heavy IMF in GCs, a prominent core of bright stars tends to emerge \citep{Chatterjee2017,Giersz2019,Weatherford2021,Wang2021}. 
Therefore, to constrain the massive end of the IMF, comparisons between dynamical models and observations of GCs are required. 

Palomar 5 (Pal~5) is among the Galactic GCs renowned for its long tidal streams and unusually low central density \citep[e.g.][]{Rockosi2002,Odenkirchen2001,Odenkirchen2002,Odenkirchen2003,Koch2004,Odenkirchen2009,Carlberg2012,Kuzma2015,Ishigaki2016,Price-Whelan2019,Bonaca2020,Starkman2020}, which suggests the possible presence of a substantial number of BHs in the cluster \citep[][hereafter G21]{Gieles2021}. 
Understanding the properties of the BH population in Pal~5 is also crucial for explaining the pronounced nature of its stream. 
G21 employed self-consistent $N$-body models that resolve individual stars to propose the existence of a large population of BHs in the cluster core (20\% of the total mass), enhancing tidal disruption. 
However, the BH hypothesis needs further confirmation, because the observed density profiles of the cluster and the stream could also be reproduced by an $N$-body model of a BH-free cluster with a low initial density.

The binary population of Pal 5 plays a crucial role in resolving this degeneracy. 
According to the \cite{Heggie1975}-\cite{Hills1975} law, close encounters with binaries can result in two opposing evolutionary trends: wide/soft binaries become less bound and decay with a few close encounters, while tight/hard binaries become tighter due to the increased kinetic energy of the intruder and the centre-of-mass of the binary. 
The boundary between these two types depends on the local kinetic energy of particles where the binary resides. 
G21 argue that the kinetic energy of BHs is higher than that of stars in a cluster without BHs with similar half-light radius. 
It is therefore expected that fewer soft binaries could survive in the case the cluster contains BHs, which is a prediction that can be tested with observations.

Furthermore, due to the large distance of Pal 5, most binaries cannot be resolved spatially by current state-of-art observational instruments. 
Because unresolved binaries might influence the determination of velocity dispersion and present-day mass functions, it is worthwhile to investigate how primordial binaries and BHs collectively affect the line-of-sight velocity measurement and mass function and whether it can be used to indirectly constrain the existence of BHs.

In this study, we perform $N$-body simulations of several Pal~5-like clusters with and without BHs and incorporating a large number of binaries, to examine the impact of BHs on binary disruption and the long-term evolution of Pal~5 and its tidal tails. 
Section~\ref{sec:method} describes the $N$-body simulation method, data analysis tools, and the observational data of Pal~5 utilized in this study. 
Section~\ref{sec:result} presents the results of our $N$-body models, comparing the structural evolution, surface number density, binary properties, and present-day mass function with models from G21 and observational data. 
Section~\ref{sec:discussion} discusses the limitations of our models and outlines prospects for future observations. 
Finally, Section~\ref{sec:conclusion} concludes this work.

\section{Methods}
\label{sec:method}

\subsection{{\it N}-body code}

We conducted simulations of Pal 5-like clusters using the high-performance $N$-body code \textsc{petar} \citep{Wang2020b}.
To achieve high parallel performance, the framework for developing parallel particle simulation codes (\textsc{fdps}) is implemented in \textsc{petar} \citep{Iwasawa2016,Iwasawa2020}. 
The code incorporates the particle-tree and particle-particle method (P$^3$T) \citep{Oshino2011}, which enables the separate integration of long-range and short-range interactions between particles. 
For accurate integration of the weak long-range interactions, the code uses a \cite{Barnes1986} particle-tree method with a 2nd-order Leap-frog integrator, which has a computational cost of $O(N\log(N))$. 
To accurately follow orbital motions of binaries, hyperbolic encounters, and the evolution of hierarchical few-body systems, the 4th-order Hermite method along with the slowdown-algorithmic regularization (SDAR) method is used \citep{Wang2020a}. 
One of the major advantages of the \textsc{petar} code is its capability to include a large fraction of binaries, up to 100\%, in the simulation of stellar systems without significant performance loss. 
This feature enables us to carry out the models presented in this work.

In our simulations, we included binaries with a wide period distribution (see Section~\ref{sec:init}), requiring the use of Leap-frog, Hermite, and SDAR integrators for integrating binary orbits. 
While Leap-frog and SDAR are symplectic methods that conserve energy and angular momentum, the Hermite integrator does not. 
We employ sufficiently small time steps for the Hermite integrator to ensure that the artificial drift of semi-major axes and eccentricities remains insignificant throughout the entire evolutionary time of all our models.
The key parameters for switching the integrator and controlling the accuracy of one simulation in this work are provided below:
\begin{itemize}
    \item Changeover inner radius: 0.0027~pc
    \item Changeover outer radius: 0.027~pc
    \item SDAR separation criterion: 0.000216~pc
    \item Tree time step: 0.0009765625~Myr
    \item Hermite time step coefficient $\eta$: 0.1
\end{itemize}
See \cite{Wang2020b} for the details on the definition of these parameters.

The population synthesis code for single and binary stellar evolution, \textsc{sse} and \textsc{bse}, are implemented in \textsc{petar} \citep{Hurley2000,Hurley2002}.
Furthermore, the code utilizes an updated version from \cite{Banerjee2020} that incorporates semi-empirical stellar wind prescriptions from \cite{Belczynski2010,Vink2011}, a "rapid" supernova model for remnant formation and material fallback from \cite{Fryer2012}, and the pulsation pair-instability supernova (PPSN) model from \citet{Belczynski2016}.
By including or excluding fallback we control the retention of BHs in our simulations.

\subsection{Milky Way potential}

The Milky Way potential is modeled by combining the \textsc{galpy} code \citep{Bovy2015} with \textsc{petar}. 
We adopt the setup of a three-component Milky Way model from G21.
The parameters are as follows:
  \[
      \begin{array}{lp{0.8\linewidth}}
         Bulge  & \cite{Hernquist1990}     \\
                & -- scale radius: 0.5 kpc                    \\
                & -- mass: $5\times 10^9 \mathrm{M}_\odot$ \\
         Disk &  \cite{Miyamoto1975} \\ 
              & -- scale length: 3.0~kpc\\
              & -- scale height: 280~pc \\
              & -- mass: $6.8 \times 10^{10} \mathrm{M}_\odot$ \\
         Halo &  \cite{NFW1996} \\
              & -- scale radius: 16~kpc \\
              & -- virial mass: $8.127 \times 10^{11}~\mathrm{M}_\odot$ \\
              & -- concentration: 15.3 \\
      \end{array}
   \]
\noindent

The present position of Pal~5 obtained in G21 is [5.733, 0.2069, 14.34] kpc and [-41.33, -111.8,-16.85] km~s$^{-1}$ in the cartesian Galactocentric frame.
The corresponding observational quantities of Pal~5 are:
\\\\
    \begin{tabular}{l|l}
        RA & 229.0217 deg \\
        Dec & -0.1109 deg \\
        Distance from Sun & 19.98 kpc \\
        Proper motion [RA~$\cos{(\mathrm{Dec}})$] & -2.67 mas~yr$^{-1}$ \\
        Proper motion [Dec] & -2.67 mas~yr$^{-1}$ \\
        Radial velocity &  -57.5 km s$^{-1}$\\
    \end{tabular}
\\\\

\begin{figure}
	\includegraphics[width=\columnwidth]{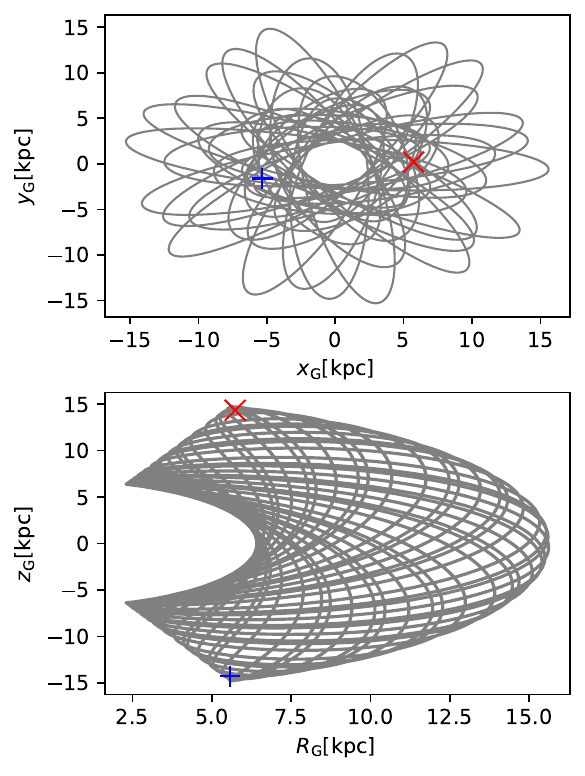}
    \caption{The orbit of the Pal~5 in the Galactocentric frame. The upper and lower panels show the projected trajectory in the $x_\mathrm{G}-y_\mathrm{G}$ and $R_\mathrm{G}-z_\mathrm{G}$ planes, respectively. $R_\mathrm{G}$ is the projected radial coordinate in the $x_\mathrm{G}-y_\mathrm{G}$ plane. The symbols '+' and 'x' represent the zero-age and present-day positions, respectively.}
    \label{fig:orbit}
\end{figure}

G21 has derived the initial position and velocity of Pal~5 ($\sim$11.5~Gyr ago) by backward integrating the orbit. 
But due to the different implementation of Galactic  potentials in \textsc{nbody6} used in G21 and in \textsc{galpy}, we could not directly use it.
Using \textsc{galpy}, we trace back the orbital motion of Pal~5 in a similar way and obtain the initial position and velocity as $[-5.339, -1.602, -14.27]$ kpc and $[-21.78, 111.9, -45.52]$ km~s$^{-1}$, respectively.
The orbit of Pal~5 calculated by \textsc{galpy} is shown in Figure~\ref{fig:orbit}.

\subsection{Mock photometry}

To convert snapshots from the $N$-body models to photometric data for different filters used in observations, we use the code \textsc{galevnb} \citep{Pang2016RAA}, which selects corresponding spectral templates from the library of \citet{Lejeune1997,lejeune1998} according to the fundamental stellar properties, such as stellar mass, temperature, luminosity and metallicity from $N$-body simulations. 
By convolving the spectra with the filter response curve from a given filter, we obtain the observational magnitudes of specific filters of main-stream telescopes, such as Hubble Space Telescope (HST) and the future Chinese Survey Space Telescope (CSST) for individual stars in the $N$-body models. 
In this way, we produce mock observations for $N$-body models, which allows a direction comparison with observational data.
This is useful to compare the density or surface brightness profiles, unresolved binaries and stellar mass functions between observations and the models.

In this study, the line-of-sight velocity of unresolved binaries is calculated using the Johnson I-band filter (as described in Section~\ref{sec:dvr}).
For creating the color-magnitude diagram, we employ the HST F555W and F814W filters, along with the CSST g and i filters.
To convert luminosity to mass for unresolved binaries, we utilize the HST F555W filter. Further details can be found in Section~\ref{sec:mf}.

\subsection{Observational data}


To validate our $N$-body model and ensure its accuracy in reproducing the surface number density $\Sigma(R)$ and mass function of Pal~5, we compare it with observational data. We utilize the data from \cite{Ibata2017} for the surface number density and the masses of stars obtained from two HST observations with Program IDs 6788 \citep[PI: Smith;][]{Grillmair2001} and 14535 (PI: Kuepper) as reported in \cite{Baumgardt2023}.

The observed surface number density $\Sigma(R)$ encompasses stars with g-band magnitudes ranging from 19 to 23, with photometry obtained from the Canada-France-Hawaii Telescope. The corresponding mass range of these stars is 0.625 to 0.815 $M_\odot$, determined using the magnitude-mass conversion provided by G21.

Regarding the masses of stars derived from the HST data, \cite{Baumgardt2023} employed Dartmouth isochrones to fit the CMDs of the clusters and employed them to convert magnitudes into masses. Further details can be found in their work.

\subsection{Star cluster models}
\label{sec:init}

To reproduce Pal~5's observed surface density and present-day position in the Galaxy, we generate the initial conditions of $N$-body models by referring to the wBH-1 and noBH-1 models in G21, which have the closest property to the observational data assuming Pal~5 contains a cluster of BHs and no BH, respectively.

For the wBH-1 model, natal kick velocities of BHs after supernovae are affected by the material fallback from \cite{Fryer2012}. 
A large fraction of BHs are retained in the clusters and finally sink to the centre via dynamical friction.
The existence of a BH subsystem can significantly affect the structure and evolution of star clusters.
As a result, the cluster has a loose core of luminous stars.
The wBH-1 model has an initial half-mass radius, $\rhz = 5.85$~pc, and an initial number of stars, $N_0=2.1\times10^5$.

In contrast, the noBH-1 model assumes BHs have the same high kick velocities as neutron stars and almost none are retained after supernova explosions.
Without BHs, the core collapse of luminous stars result in a dense core. 
In order to reproduce the observed surface brightness profile, G21 find that the cluster must therefore have had a much lower density initially.
Thus, for the noBH-1 model, $\rhz=14$~pc and $N_0=3.5\times10^5$.

We conducted five $N$-body models with varying setups of primordial binaries and the presence of BHs. 
The initial conditions for these five models are summarized in Table~\ref{tab:init}. 
We assigned labels to the models to indicate the existence of primordial binaries and BHs.

For BH treatment, models with the label "BH" refer to the wBH-1 model from G21, where the mass fallback scaling for kick velocities is applied so that a part of the BHs has low kick velocities and stays in the clusters. 
They also have the same $N_0$ and $\rhz$ as those of the noBH-1 model.

Models with the label "noBH" refer to the noBH-1 model from G21.
In these models, all BHs have high kick velocities similar to the neutron stars after asymmetric supernovae.
The velocity distribution follows a (1D) Maxwellian distribution with a dispersion of 265 km/s.
As a result, we found no BHs are retained in our noBH models.

The prefix "noBin" and "Bin" represent without and with primordial binaries, respectively.
For "Bin" models, all stars are in binaries initially.
For massive binaries with the component mass $>5\,{\rm M}_\odot$, except the Bin-noBH-F model, all other "Bin" models have the period and mass ratio distributions follow the observational constraints of OB binaries from \cite{Sana2012}.

For low-mass binaries, except the Bin-BH-Alt model, all other "Bin" models assume the properties of primordial binaries following the model from \cite{Kroupa1995a,Kroupa1995b} and \cite{Belloni2017} (naming as Kroupa binary model). 
The orbital parameters of this model are derived from the inverse dynamical population synthesis of binaries in the Galactic field.
This model assumes an universal property of primordial binaries and all stars forming in star clusters.
In addition, a correction of the period and eccentricity distributions from \cite{Belloni2017} is included to better fit the observational data of GCs.

\begin{figure}
    \includegraphics[width=\columnwidth]{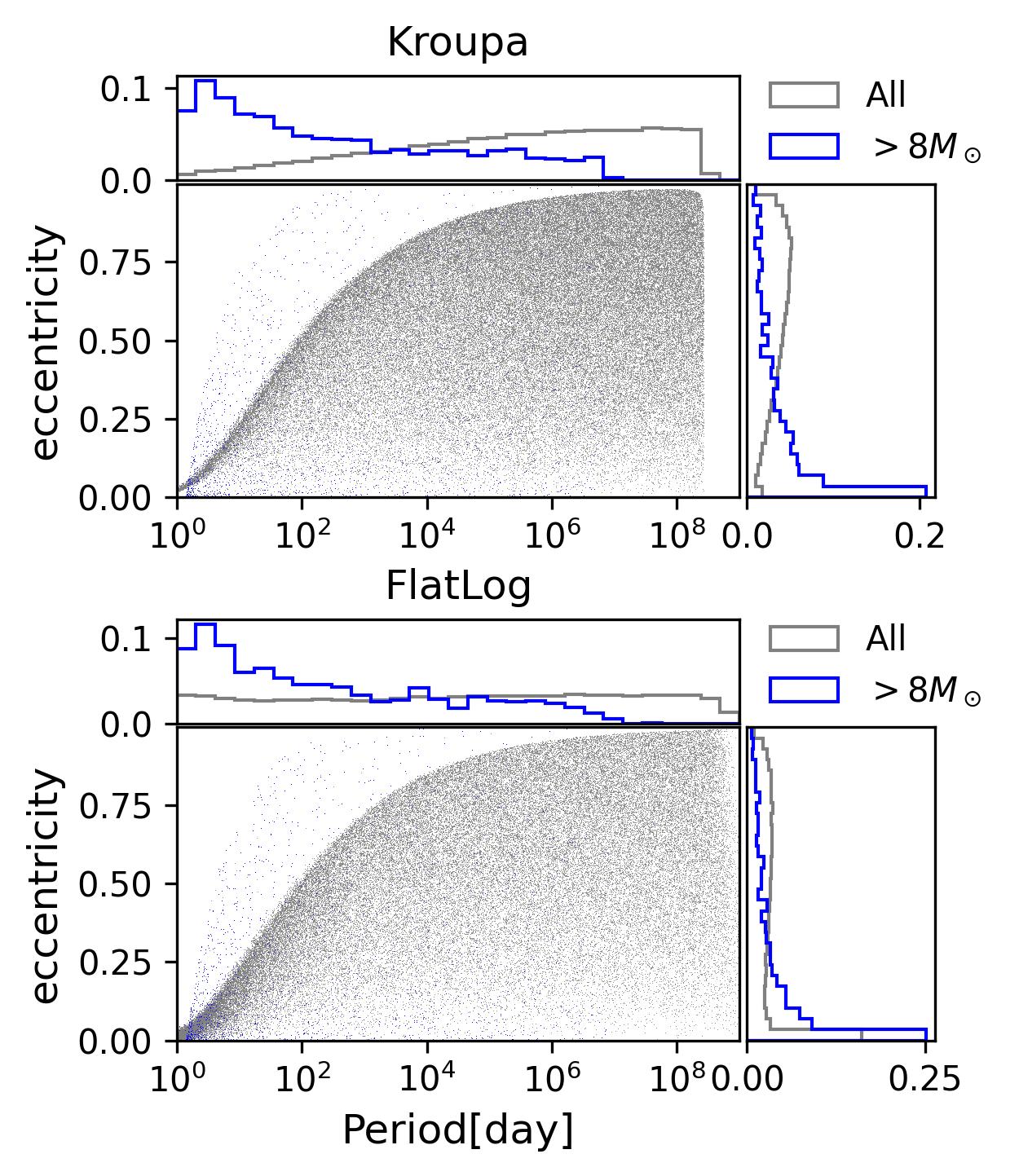}
    \caption{Initial periods ($P$) v.s. eccentricities ($e$) of primordial binaries for the Kroupa binary model and the FlatLog model. The central plot of each panel shows $P$-$e$ of individual binaries. The upper and the right histograms show the normalized distribution of $P$ and $e$, respectively.
    The distribution of massive binaries is shown by blue lines. }
    \label{fig:b0ae}
\end{figure}

For the Bin-BH-Alt model, we assume a different setup of low-mass primordial binaries (referred to as FlatLog model) as a comparison with the Kroupa binary model.
The semi-major axes follow a flat distribution in the logarithmic scale where the minimum and maximum value are $3$ solar radius and 2 pc, respectively.
The eccentricity and mass ratio distributions are the same as those of the Kroupa binary model.

The period and eccentricity distributions are shown in Figure~\ref{fig:b0ae}.
For both binary models, the initial distribution of periods covers a wide region with 9 orders of magnitudes. 
The initial eccentricities exhibit a sharp peak at $e=0$ and a broader peak at $e=0.8$, respectively.
All binaries with peri-centre separation less than the sum of the stellar radii of the two components are excluded. 
Thus, an empty region is visible in the period-eccentricity distribution of Figure~\ref{fig:b0ae}.
In addition, the eccentricity distributions of the Kroupa and FlatLog are different after adjustment. 

These binary setups cover a wide range of binary orbital periods, where a large fraction of binaries are unstable in the cluster environment. 
After a short time (about one crossing time), the binary fraction significantly reduces.
Referring to Pal~5, the binary fraction of our setup may be overestimated.
The benefit is that we can investigate how long-term dynamical evolution of the clusters with and without BHs affect both the tight and wide binaries.

The Bin-noBH-F model has the same $N_0$ and $\rhz$ as those in the noBH-1 model.
However, after finishing the simulation, we found that the Bin-noBH-F model cannot reproduce the final structure of the noBH-1 model at 11.5~Gyr and it has sufferred complete tidal disruption before 10 Gyr.
The suffix "F" in the name of the model indicates that this is a failed model.
Thus, we conducted another model "Bin-noBH" by reducing $\rhz$ to $13.2$~pc.
This small modification results in a cluster similar to Pal 5 after 11.5~Gyr.

In addition, we excluded massive binaries in the Bin-noBH-F model to prevent non-supernovae BH formation in a binary, but we observed that such events did not occur. 
Therefore, in the Bin-noBH model, we added the Sana distribution to massive binaries to ensure consistency with the Bin-BH models.

The common setup for all models is also summarized in Table~\ref{tab:init}. 
All models were evolved for a duration of $12.0$ Gyr.
At 11.5~Gyr, the clusters are located at the same Galactic position as Pal~5. 
However, since the model did not precisely reproduce the surface number density of Pal5, we continue to evolve the cluster further to determine the age (referred to as $\Tmat$) when the model matches the observation more closely, as detailed in Section~\ref{sec:ndp}.
We assumed a spherically symmetric Plummer profile \citep{Plummer1911}  with no primordial mass segregation. 
The initial mass function (IMF) of stars followed the two-component power-law shape described by \cite{Kroupa2001}. 
We adopted the same mass range of $0.1-100M_\odot$ as used in G21, and the power-law indices ($\alpha$) and mass ranges are described as:
\begin{equation}
\label{eq:imf}
\alpha = \left \{\begin{matrix}
-1.3 &  (0.1<m<0.5~M_\odot)\\
-2.3 &  (0.5<m<100~M_\odot) 
\end{matrix} \right .
\end{equation}

In this study, we adopted a cluster metallicity of $Z=0.0006$, which is consistent with the value reported in \cite{Smith2002} of $[\mathrm{Fe/H}] \approx -1.4$ dex for Pal~5. 
The initial star cluster models were generated using the updated version \citep{Wang2019} of the \textsc{mcluster} code \citep{Kuepper2011}. This update includes the implementation of the Kroupa binary model generator, as shown in Figure~\ref{fig:b0ae}.

\begin{table*}
	\centering
	\caption{Initial conditions of the $N$-body models. All models include the Plummer (1911) profile, the Kroupa (2001) initial mass function with a mass range from $0.1$ to $100~M_\odot$, a metallicity of $z=0.0006$, and a simulation duration of 12~Gyr.}
	\label{tab:init}
	\begin{tabular}{lccccc} 
		\hline
		Models & noBin-BH & Bin-BH & Bin-BH-Alt & Bin-noBH & Bin-noBH-F\\
		\hline
		$\rhz$ [pc] & 5.85 & 5.85 & 5.85 & 13.2 & 14\\
		$N_0$ & 210000 & 210000 & 210000 & 350000 & 350000\\
		Binary fraction & no & $100\%$ & $100\%$ & $100\%$ & $100\%$ \\
            Low-mass binary & no & Kroupa & FlatLog & Kroupa & Kroupa \\
            massive binary & no & Sana & Sana & Sana & no \\ 
            Retaining BH & fallback-scale & fallback-scale & fallback-scale & no & no \\
            $\Tmat$  [Gyr] & 11.8 & 12.0 & 11.0 & 12.0 & \\
		\hline
	\end{tabular}
\end{table*}

\section{Results}
\label{sec:result}

\subsection{Structural evolution}

First, we present the evolution of the cluster structure and compare our results to the models from G21 and the observational data.
Generally, although the existence of binaries does not significantly affect the structural evolution, the small difference can be amplified by the Galactic tidal field and result in early dissolution of the Bin-noBH-F model.
In addition, the existence of primordial binaries reduces the BH populations and results in shorter relaxation times in the early evolution. 
The stochastic formation of BBHs also affects the expansion of the cluster and eventually influences the disruption of the cluster.
The surface number density of $N$-body models roughly agree with observations with a larger central density.

\subsubsection{Half-mass relaxation time}
\label{sec:trh}

The two-body relaxation time is an important timescale of stellar dynamics, which reflects the speed of changes in the density and mass segregation of a cluster and its tidal dissolution.
The one-component half-mass relaxation time ($\trha$) defined in \cite{Spitzer1987} has the form as
\begin{equation}
    \trha = 0.138 \frac{N^{1/2} \rh^{3/2}}{m^{1/2} G^{1/2} \ln \Lambda},
    \label{eq:trh}
\end{equation}
where $N$ is number of stars, $\rh$ is the half-mass radius, $m$ is the average mass of stars, $G$ is the gravitational constant, and $\ln \Lambda$ is the Comlumb logarithm.
When BHs exist, the binary heating is dominated by BBHs, $\trha$ leads to an underestimation of the relaxation timescale of the system. 
\cite{Wang2020} found that a proper two-component relaxation time ($\trh$) can be obtained by dividing a correction factor $\psi$, defined as
\begin{equation}
    \psi = \frac{n_1 m_1^2 / \sigma_1 + n_2 m_2^2 / \sigma_2 }{n m^2 / \sigma},
\end{equation}
and 
\begin{equation}
    \trh = \frac{\trha}{\psi}
\end{equation}
where the suffixes 1 and 2 represent the quantities for non-BH and BH components, respectively.

\begin{figure}
    \centering
    \includegraphics[width=\columnwidth]{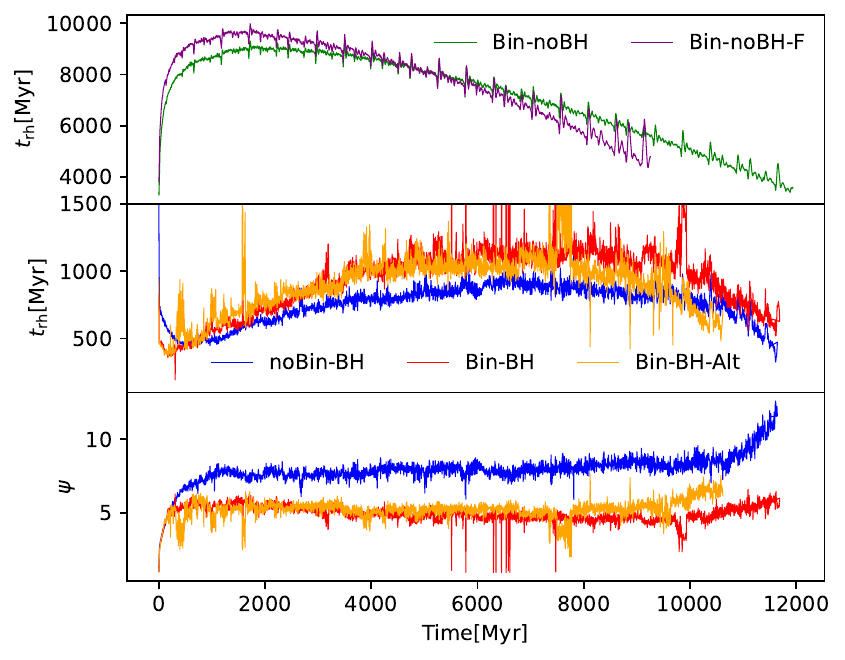}
    \caption{The evolution of two-component half-mass relaxation time for all models ($\trh$; upper two panels) and $\psi$ factors (lower panel) for BH models. }
    \label{fig:trh}
\end{figure}

Figure~\ref{fig:trh} illustrates the evolution of $\trh$ and $\psi$. 
The three BH models exhibit significantly shorter $\trh$ compared to the noBH models. 
During the first 100 Myr, the noBin-BH model displays a longer $\trh$ compared to the Bin-BH and Bin-BH-Alt models because the Bin models treat binaries as single objects when calculating $\trh$.
Consequently, the Bin-BH and Bin-BH-Alt models experience relatively faster expansion of $\rh$ and faster mass segregation of BHs (see Section~\ref{sec:rh}). 
Subsequently, the trend reverses, and the $\trh$ of the noBin-BH model becomes shorter than that of the Bin-BH and Bin-BH-Alt models due to the difference in the number of BHs (see Section~\ref{sec:massloss}). 
As a result, the $\rh$ of the noBin-BH model expands faster than that of the other two models.
After 8 Gyr, the $\trh$ of all three BH models starts to decrease due to mass loss via tidal evaporation.

The values of $\psi$ for the BH models exceed 5, indicating that BHs significantly impact the relaxation process of the clusters. 
Further discussion of $\rh$ is provided in Section~\ref{sec:rh}.

In contrast, the two noBH models exhibit much longer $\trh$. 
There is a rapid increase in $\trh$ during the first 100 Myr, primarily due to the strong stellar winds from massive stars and the escape of BHs. 
Consequently, although the morphology appears similar at 11.5 Gyr for models with and without BHs, the relaxation processes differ significantly. 
These differences can lead to variations in the properties of binaries. 
In Section~\ref{sec:binary}, we analyze the impact of these differences and discuss their implications for binary systems. 
It is important to note that assuming $\psi=1$ for the noBH models is not accurate, as there is still an order of magnitude difference between the minimum and maximum masses of stars.


\subsubsection{Half-mass radius}
\label{sec:rh}

\begin{figure}	
    \includegraphics[width=\columnwidth]{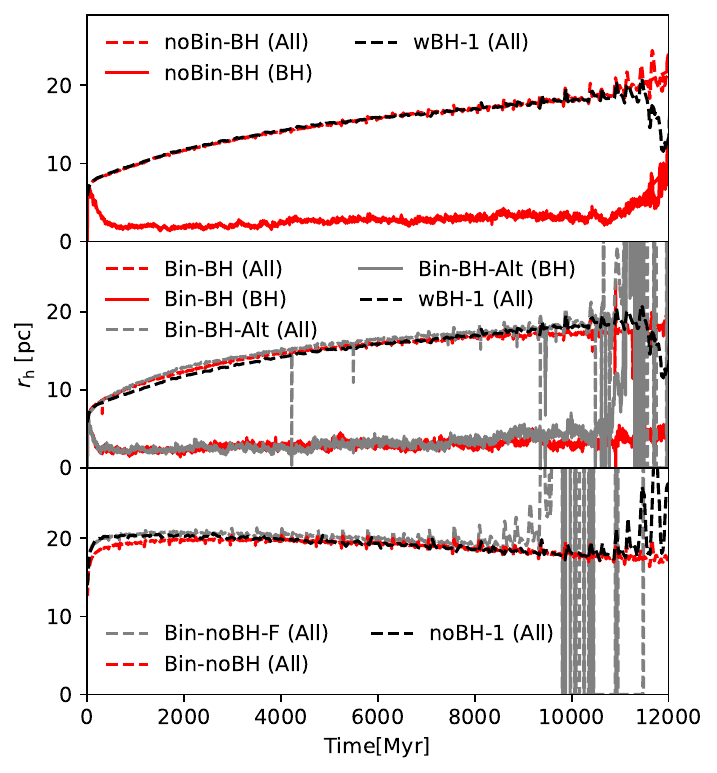}
    \caption{The evolution of half-mass radius of all objects ($\rh$; dashed curves) and the half-mass radius of BHs ($\rhbh$; solid curves). The wBH-1 and noBH-1 models from Gieles et al. (2021) are shown as references.}
    \label{fig:rht}
\end{figure}

Figure~\ref{fig:rht} illustrates the evolution of $\rh$ for all models, including the ones from G21 for comparison. 
We observe that the presence of primordial binaries has a weak impact on the evolution of $\rh$, consistent with the theoretical findings of \cite{Wang2022}. 
When BHs exist, the long-term structural evolution of star clusters is primarily controlled by binary heating driven by the dynamical interactions between BBHs and the surrounding objects at the cluster center. 
The majority of primordial binaries have much smaller masses compared to BBHs, and therefore have a negligible impact on the binary heating until most BHs have escaped from the cluster. 
A small subset of massive primordial binaries can eventually evolve into BBHs. 
However, even in the absence of these massive binaries, a star cluster can generate BBHs through chaotic three-body interactions when the central density of the cluster reaches a threshold after the core collapse of BHs (see Section~\ref{sec:bh}). 
Consequently, we only observe minor differences of $\rh$ between the Bin-BH, Bin-BH-Alt, and wBH-1 models during the first 10 Gyr of evolution. 
This can be explained by the differences in relaxation times ($\trh$) discussed in Section~\ref{sec:trh}.
The galactic potential also affects $\rh$, but since all models share the same orbit, the influence is similar.

However, after 10 Gyr, the Bin-BH-Alt model exhibits a similar $\rh$ to that of the wBH-1 model, but its $\rh$ shows significant variations, indicating an energy imbalance and the onset of a disruptive tidal phase. 
In contrast, both the Bin-BH and wBH-1 models remain stable until 12 Gyr. 
This differing behavior is attributed to stochastic BBH heating, as explained in Section~\ref{sec:bh}.

The BH models with binaries (Bin-BH) and without binaries (noBin-BH) exhibit different timescales for the mass segregation of black holes, as indicated by the initial rapid contraction of $\rhbh$. 
In the Bin-BH model, $\rhbh$ undergoes faster contraction during the early stages of evolution compared to the noBin-BH model. 
This disparity can be attributed to the difference in $\trh$, as the timescale for mass segregation is proportional to $\trh$.

When comparing the noBH models with binaries (Bin-noBH-F) and the model from G21 without binaries (noBH-1), significant differences in the evolution of $\rh$ emerge after 8 Gyr. 
The Bin-noBH-F model experiences tidal disruption at around 9 Gyr, whereas the noBH-1 model survives until 11.5 Gyr.  G21 noted that the final properties of the noBH models are more sensitive to changes in the initial conditions, and in fact argued that this `fine tuning' problem disfavours the noBH scenario.
An offset of $\rhz$ needs to be introduced in the Bin-noBH model to achieve consistent $\rh$ at 11.5 Gyr.

Two factors may explain the need for this offset. 
Firstly, in the absence of BHs, binary heating is primarily generated by low-mass binaries. 
Consequently, the influence of primordial binaries is more pronounced compared to models with BHs. 
Secondly, due to the larger $\rhz$, the cluster becomes more sensitive to the galactic tide. 
The presence of primordial binaries affects the relaxation time of the system, as the dynamical effect of tight binaries is equivalent to that of single objects, resulting in a shorter relaxation time for the system. 
Consequently, the system dissolves faster, necessitating a denser initial cluster to allow the cluster's survival, as seen in the noBH-1 model.
Additionally, the differences caused by the stochastic scatter of $\rh$ resulting from the random seeds used to generate the initial conditions may also be amplified by the galactic tide, contributing to the divergent evolution.

\subsubsection{Mass loss}
\label{sec:massloss}

\begin{figure*}
	\includegraphics[width=\textwidth]{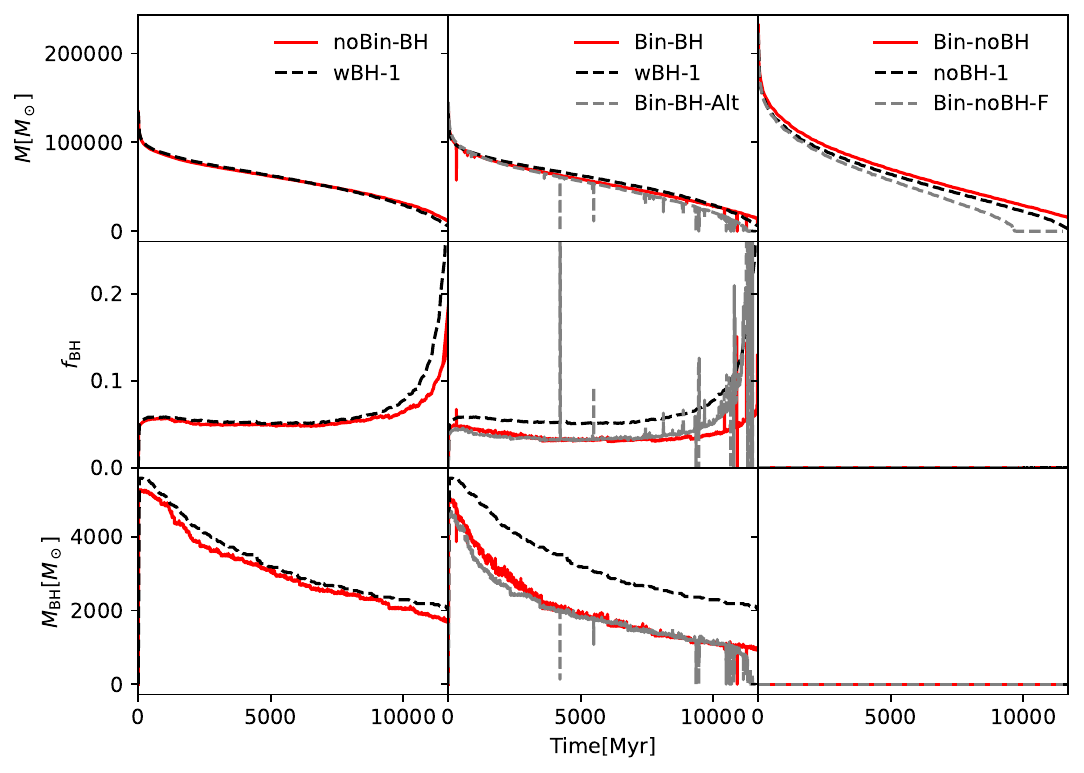}
        \caption{The evolution of the bound mass  ($M$), the BH mass fraction ($\fbh$) and the bound mass of BHs ($\Mbh$) for all models. The data of the wBH-1 and noBH-1 models are shown for comparison.}
    \label{fig:Mfbh}
\end{figure*}

The upper panels of Figure~\ref{fig:Mfbh} show the evolution of the total mass ($M(t)$) of our models. 
Data of the wBH-1 and the noBH-1 from G21 are also shown as references.
The mass loss has two channels: wind mass loss driven by stellar evolution and escapers via stellar dynamics of star clusters.
To have a consistent definition of $M$, all models use the same criterion to select escapers.
First, we calculate the bound energy of stars and centre-of-the-mass of binaries without external potential and then select escapers with energy >0.

Here we compare the three cases:
For models with no primordial binary and with BHs, $M(t)$ of our noBin-BH model agrees with the wBH-1 model from G21.
The final mass of the noBin-BH model at 11.5~Gyr is slightly larger than that of the wBH-1 model.

For models with primordial binaries and with BHs, compared to the wBH-1 model, the Bin-BH and the Bin-BH-Alt models lose mass faster during the first few hundred Myr, but mass loss of the Bin-BH model becomes slower near the end of the simulation.
Finally, the Bin-BH and the wBH-1 models agree with each other, while the Bin-BH-Alt model dissolves after about 11~Gyr.

For models with no BHs, the Bin-noBH-F model with primordial binaries loses mass faster than the noBH-1 model with no binaries.
The Bin-noBH model, with a smaller $\rhz$, experiences a relatively slower mass loss, and its $M(t)$ remains slightly above that of the noBH-1 model at 11.5~Gyr.
In general, the evolution of $M(t)$ and $\rh$ are similar for all three cases.


\subsubsection{Black holes}
\label{sec:bh}

BHs significantly affect the long-term dynamical evolution.
We investigate the mass fraction of BHs $(\fbh)$ and the bound mass of BHs ($\Mbh$) in Figure~\ref{fig:Mfbh}.
The evolution of $\fbh$ in the noBin-BH and the wBH-1 models agree with each other in the first 8~Gyr.
Then, $\fbh$ increases more slowly in the noBin-BH model and is half that in the wBH-1 model at 11.5~Gyr.
$\Mbh$ of the noBin-BH model is slightly smaller than that of the wBH-1 model initially and such a difference is inherited in the long-term evolution.
Finally, as a large fraction of stars escape, such initial differences lead to a large difference of $\fbh$ at the end.

For the Bin-BH and the Bin-BH-Alt models, $\Mbh$ are significantly smaller than that of the noBin-BH model during the early evolution.
This difference is due to the stellar evolution of massive binaries. 
Based on the orbital parameters of binaries from \cite{Sana2012}, the progenitors of BHs (massive stars) are all in binaries. 
A fraction of the tight binaries suffers mass transfer and mergers. 
The BHs formed from these binaries can have different distribution of masses.
The maximum $\Mbh$ of the Bin-BH model is about $250~M_\odot$ less than that of the noBin-BH model.
Then, after the mass segregation of BHs (a few hundreds Myr), binary heating of BBHs start to kick out BHs from the cluster, and result in larger difference of $\Mbh$ during the long-term evolution. 
Although the Bin-BH (Bin-BH-Alt) and the noBin-BH models show a large difference of $\Mbh$, their evolution of $M$ and $\rh$ is similar before 10~Gyr.
This was also observed in \cite{Wang2022}.

\begin{figure}	
    \includegraphics[width=\columnwidth]{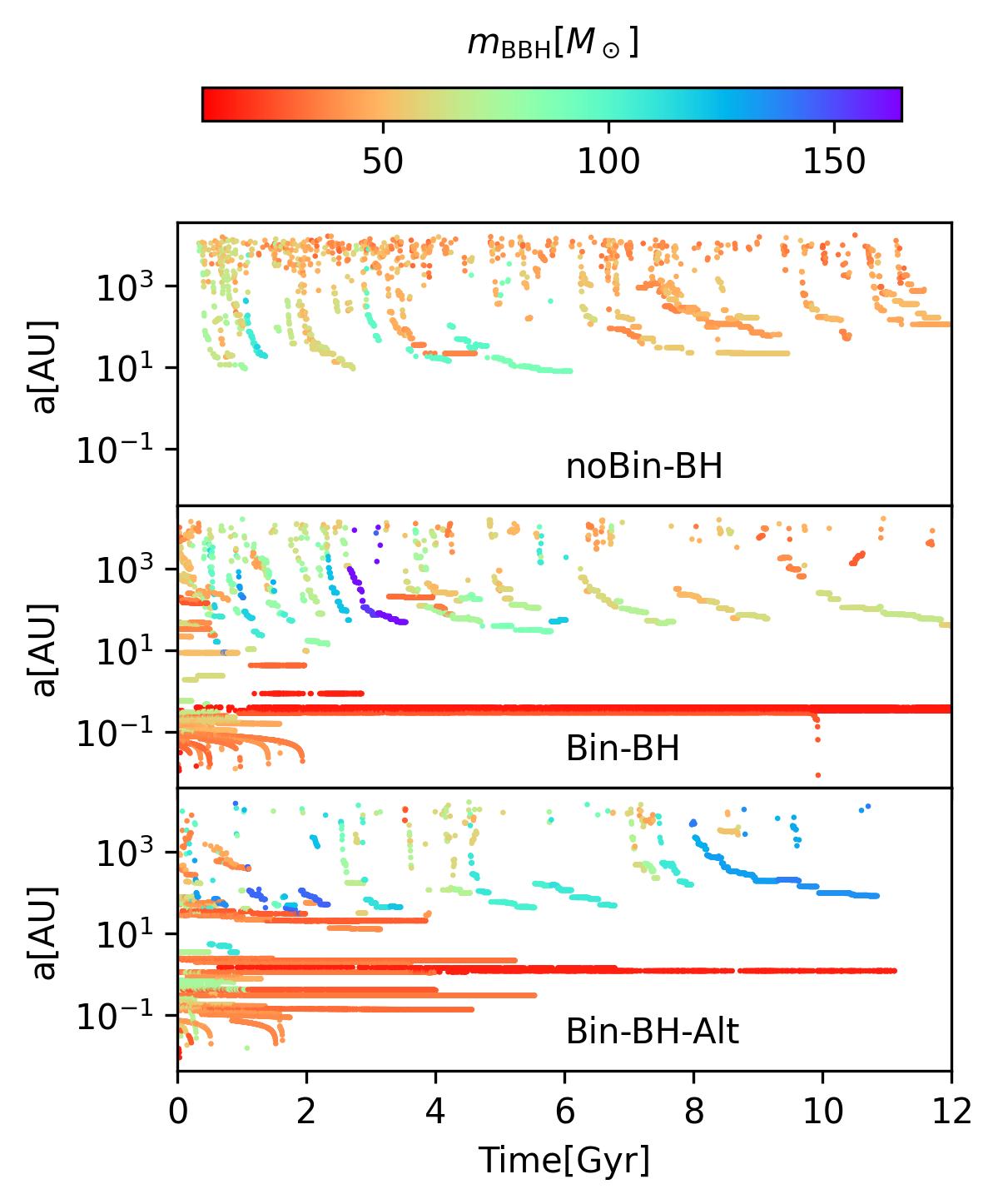}
    \caption{The evolution of the semi-major axes of BBHs within the core radius ($\rc$) of the three BH models. The colors of the lines indicate the masses of the BBHs. We can observe a reduction of the semi-major axes of individual BBHs, indicating their dynamical hardening over time ($a>10$~AU) and inspiral by GW radiation ($a<1$~AU).}
    \label{fig:abh}
\end{figure}

The evolution of the semi-major axes ($a$) of BBHs reflects both binary heating and mergers driven by gravitational wave (GW) radiation.
Figure~\ref{fig:abh} provides a comparison of this evolution for the three BH models. 
Despite the absence of primordial binaries in the noBin-BH model, we can still observe the formation of BBHs and their orbital contraction. 
The frequency of BBH formation and the overall trend of $a$ are similar for all three models, except that the two models with primordial binaries exhibit a higher number of BBHs formed from these binaries during the first 1000 Gyr.
Some of these BBHs with $a<1$ AU undergo orbital shrinking due to GW radiation, ultimately merging to form more massive BHs. 
These newly formed BHs lead to the creation of massive BBHs with masses exceeding 100 $M_\odot$.
The presence of these massive BBHs can have a substantial impact on the evolution of the star cluster, influencing its dynamical and structural properties.

In particular, for the Bin-BH-Alt model, the formation of a massive BBH around 8 Gyr coincides with a faster expansion of $\rh$ compared to the Bin-BH model, ultimately leading to an earlier disruption of the Bin-BH-Alt model. 
Hence, the divergent evolution of the Bin-BH and Bin-BH-Alt models after 8 Gyr is attributed to the stochastic formation of BBHs.

It is important to note that our models do not account for the high-velocity kicks experienced by newly formed black holes due to asymmetric GW radiation following mergers. 
Therefore, the formation of such massive BBHs might not be as common as our models suggest. 
Consequently, the stochastic effect of massive BBH heating could be overestimated in our cases.

\subsubsection{Surface number density profiles}
\label{sec:ndp}

\begin{figure}	
    \includegraphics[width=\columnwidth]{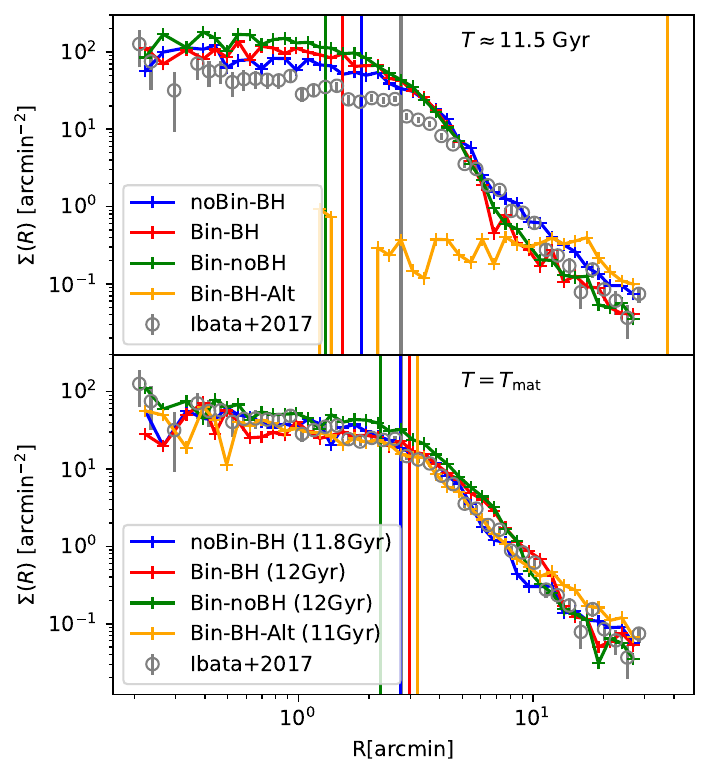}
    \caption{The surface number density ($\Sigma(R)$) profiles are presented for the $N$-body models along with observational data from Ibata et al. (2017). The upper panel displays snapshots of the $N$-body models at the present-day Galactic position and at apporximately 11.5 Gyr. The lower panel shows $N$-body snapshots that match the observed $\Sigma(R)$ profile. The ages of the corresponding snapshots ($\Tmat$) are indicated in the legend. Vertical lines are used to indicate the the `effective radius' -- the radius containing half the number of stars in projection -- ($\rhn$) of the clusters.}
    \label{fig:ndp}
\end{figure}

The determination of $\rh$ and $M$ relies on the selection criteria for identifying cluster members.
When comparing the $N$-body models with observational data from Pal~5, it is challenging to use the exact same selection criterion for both. 
A more appropriate approach is to compare the surface number density ($\Sigma(R)$), where $R$ represents the angular distance from the cluster center in the International Celestial Reference System (ICRS).

Figure~\ref{fig:ndp} illustrates the $\Sigma(R)$ profiles for our $N$-body models and the observational data of Pal~5 obtained from \cite{Ibata2017}. 
To ensure consistency with the observations, only main-sequence stars with masses ranging from $0.625 M_\odot$ to $0.815 M_\odot$ are considered in the $N$-body data (see G21 for details).

No stars are removed during the simulation, allowing for the tracking of the tidal tail evolution. 
The centre-of-mass position of the star clusters in the Galaxy at exactly 11.5 Gyr does not perfectly align with that of Pal~5. 
This is due to the long-term evolution of star cluster, where the center of the cluster drifts as a result of asymmetric mass loss due to stellar winds, supernovae, and the escape of stars.
Therefore, we select snapshots from the simulations that have the closest centre-of-mass distance to that of Pal~5 whenever a comparison is required in the subsequent analysis. 
We then correct the positions and velocities of the stars by applying the offset between the centre-of-mass of the $N$-body models and the observational data. 
The results of this correction are presented in the upper panel of Figure~\ref{fig:ndp}.
Due to the complete disruption of the Bin-noBH-F model, it is not possible to determine the centre-of-mass position for this particular model.
Therefore, it is excluded from some analysis and comparisons.

The vertical lines in Figure~\ref{fig:ndp}, representing the half surface number radii ($\rhn$), indicate that all models except the Bin-BH-Alt model are more centrally concentrated than the observed Pal~5. 
In Figure~\ref{fig:Mfbh}, it is shown that these models retain more mass at 11.5 Gyr compared to the models presented in G21.

The Bin-noBH and Bin-BH models exhibit similar $\Sigma(R)$ profiles, but this similarity is coincidental since they had different initial density profiles and evolved in opposite ways, as demonstrated in Figure~\ref{fig:rht}.

Given the time-consuming nature of the simulations, it is challenging to precisely reproduce the models of G21 and the observational data. 
To enhance the comparison with the observational data, we selected snapshots at different ages that match the observed $\Sigma(R)$ profile. 
These results are displayed in the bottom panel of Figure~\ref{fig:ndp}. 
Although the tidal streams differ substantially, we can still compare the internal properties of binaries and mass functions using these snapshots.

\subsection{Binaries}
\label{sec:binary}

\subsubsection{Binding energy of binaries}

\begin{figure*}
    \includegraphics[width=\textwidth]{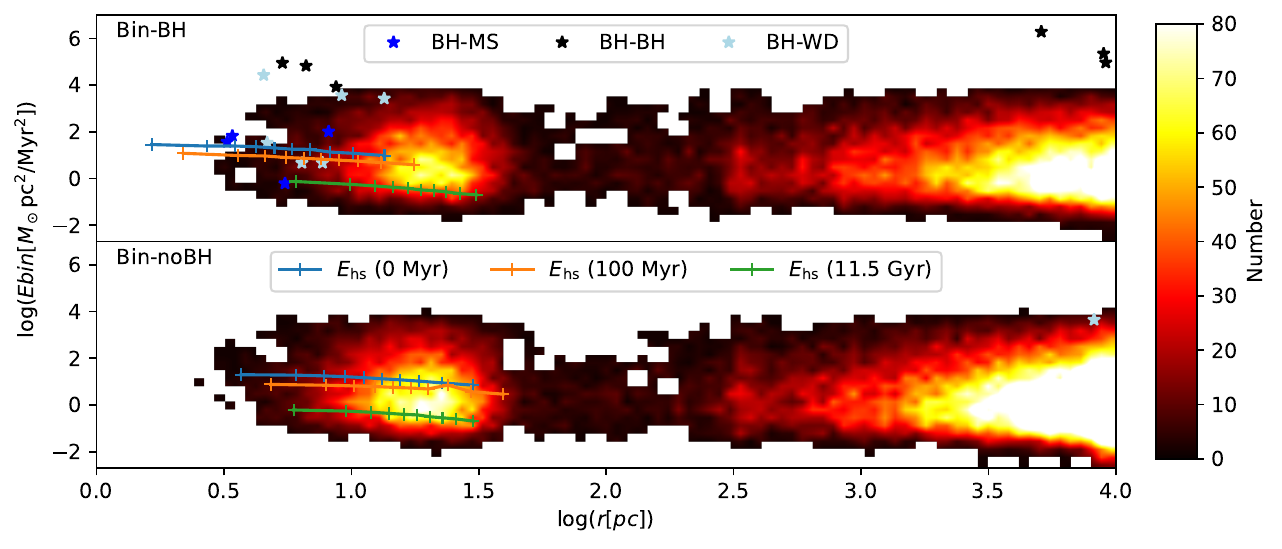}
    \caption{The contour of $r$-$\Ebin$ at $11.5$~Gyr for the Bin-BH model (upper panel) and the Bin-noBH model (lower panel). Binaries with one or two compact objects are excluded in the contour. Instead, BH-MS and BH-WD binaries are marked as blue and lightblue stars, respectively. Three curves show the hard-soft boundaries $\Ehs(r)$ at zero age, 100~Myr and 11.5~Gyr, respectively. The white region outside the color region indicates no binary.}
    \label{fig:ramap}
\end{figure*}

While the BH and noBH models may exhibit a similar $\Sigma(R)$ profile, as demonstrated in Figure~\ref{fig:ndp}, their relaxation processes differ. 
This discrepancy can lead to different properties of binaries at 11.5 Gyr.

In star clusters, perturbations from incoming objects can significantly alter the orbits of binaries. 
According to the \cite{Heggie1975}-\cite{Hills1975} law, wide or soft binaries are prone to disruption after experiencing a few close encounters with intruding objects. 
Conversely, tight or hard binaries tend to become even tighter after these encounters.

The hard-soft boundary of binding energy ($\Ehs$) at the distance to the cluster center ($r$) is determined by the local velocity dispersion:
\begin{equation}
    \Ehs = \frac{\langle m v^2\rangle}{3}
\end{equation}
where $0.5\langle mv^2\rangle$ is the average kinetic energy of stars and binaries at $r$, and $v$ is the velocity.

The hard-soft boundary of binaries evolves as the structure of the cluster changes over time.
Initially, during the first 100 Myr of star cluster evolution, there is a rapid reduction in the hard-soft boundary.
This is due to the expansion of $\rh$ caused by the strong stellar wind mass loss from massive stars, as shown in Figure~\ref{fig:rht}.

After 100 Myr, the evolution of $\rh$ slows down, and the hard-soft boundary, $\Ehs$, evolves more gradually. 
The Bin-BH and Bin-noBH models have different initial $\Ehs(r)$ curves as shown in Figure~\ref{fig:rht}, but their final $\Ehs(r)$ curves at 11.5 Gyr converge to a similar shape. 
This indicates that the distribution of binary binding energy at 11.5 Gyr may reflect the different evolutionary histories of $\Ehs$.

To further analyze the distribution of binary binding energy, Figure~\ref{fig:ramap} presents a comparison of the contour plot of $\Ebin$ versus $r$ at approximately 11.5 Gyr for the Bin-BH and Bin-noBH models. 
Across a wide range of $r$ values, spanning from the center of the cluster to the distant tidal tail, two distinct peaks can be observed. 
The first peak, located around 10-30 pc, represents the population of binaries inside the cluster. 
The second peak, with $r > 3000$ pc, corresponds to binaries that have escaped from the cluster and are distributed along the tidal tail. 

We focus on the discussion of binaries within the cluster and examine the hard-soft boundaries, $\Ehs(r)$, at three different ages: 0 Myr, 100 Myr, and 11.5 Gyr. 
These boundaries are plotted as reference curves.
To calculate $\Ehs(r)$, we divide the cluster into 10 radial bins, ensuring an equal number of objects per bin. 
Binaries are treated as unresolved objects in this analysis. 
The maximum value of $r$ is set to be at $90\%$ of the Lagrangian radius, providing a radial range that reflects the cluster's size at the three ages.

The results show that $\Ehs(r)$ does not exhibit strong variations along $r$. 
The two models, Bin-BH and Bin-noBH, have similar $\Ehs(r)$ curves, except for an offset in the radial region at 0 Myr and 100 Myr. 
The peak of $\Ebin$ falls between the $\Ehs(r)$ curves at 100 Myr and 11.5 Gyr. 
This suggests that during the first 100 Myr, not all soft binaries with $\Ebin<\Ehs$ are immediately disrupted, and many of them can survive and become hard binaries by 11.5 Gyr.

Therefore, the final distribution of $\Ebin$ does not clearly reflect the initial conditions of the two models, as anticipated by G21. 
However, the Bin-noBH model has a relatively larger number of binaries compared to the Bin-BH model. 
This difference suggests that the overall rate of binary disruption depends on the evolutionary history of the cluster density.


\subsubsection{Period distribution}
\label{sec:period}

\begin{figure}
    \includegraphics[width=\columnwidth]{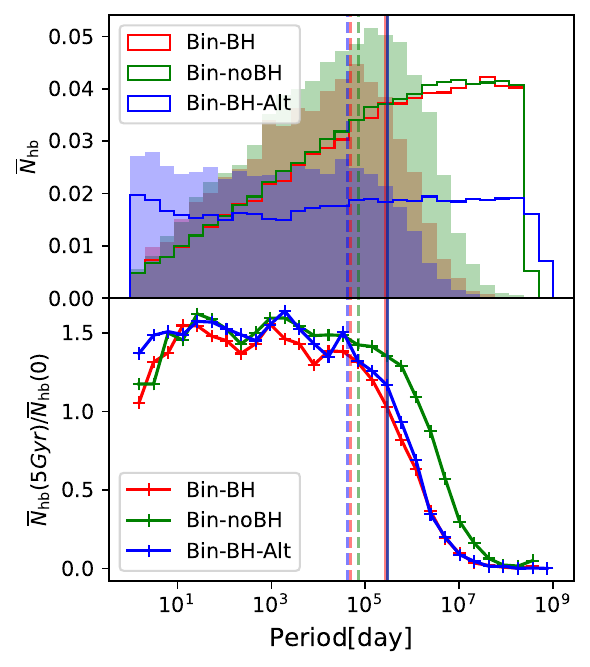}
    \caption{The period distribution of binaries within $\rh$ at two different stages: the initial phase (represented by steps) and at 5~Gyr (shown as filled histograms). 
    The upper panel displays the number of binaries within $\rh$ normalized by the bound mass of the cluster ($\nhb$) .
    The lower panel shows the ratio between $\nhb$ at 5~Gyr and the initial $\nhb$.
    The vertical dashed and solid lines represent the hard-soft boundary of period within $\rh$ at 0 and 5 Gyr, respectively.}
    \label{fig:bp}
\end{figure}

To analyze the binary disruption rate in relation to cluster dynamics, we examine the period distributions normalized by the bound mass of the cluster ($\nhb$) for three models: Bin-BH, Bin-noBH, and Bin-BH-Alt, as depicted in Figure~\ref{fig:bp}.
The period distributions at the initial phase (0 Gyr) and the median age (5 Gyr) are compared.

In the Bin-BH and Bin-noBH models, the initial period distributions are the same, but they exhibit different density profiles.
At 5 Gyr, the Bin-noBH model retains more wide binaries compared to the Bin-BH model.
The hard-soft boundaries of periods, estimated for stars within $\rh$, do not exhibit significant differences between the two models.
However, the peak of the period distribution in the Bin-BH model is closer to the hard-soft boundary at zero age, whereas in the Bin-noBH model, it aligns with the boundary at 5 Gyr.
This disparity suggests that the disruption rate of binaries is not solely determined by the hard-soft boundary.
During long-term evolution, the Bin-BH model, which is denser and contains BH subsystems, experiences a higher rate of disruption for wide binaries, resulting in the peak of the period distribution being closer to the boundary.
In contrast, the Bin-noBH model preserves more wide binaries, and the peak of the period distribution reflects the boundary at 5 Gyr for the cluster.

Comparing the Bin-BH and Bin-BH-Alt models, they share a similar density evolution but differ in the assumptions of their primordial binaries.
The ratio of $\nhb$ at 5 Gyr to the initial phase, $\nhb(5~\mathrm{Gyr})/\nhb(0)$, exhibits an identical trend for both models.
This finding implies that the binary disruption is not highly sensitive to the assumption of the initial period distribution.
Consequently, it is possible to infer the initial binary properties through inverse derivation if the evolution history of the cluster density is known (see \citealt{Kroupa1995a, Marks2011, Marks2012}).
Moreover, by utilizing the derived ratio, we can extrapolate the evolution of the period distribution of binaries for any arbitrary assumption regarding the primordial binary populations.
This provides a valuable tool for understanding the long-term dynamical evolution of binary systems within star clusters and can aid in studying the impact of different initial binary properties on the binary disruption rate and cluster dynamics.

\subsubsection{Radial distribution}
\label{sec:rdist}

\begin{figure}
\includegraphics[width=\columnwidth]{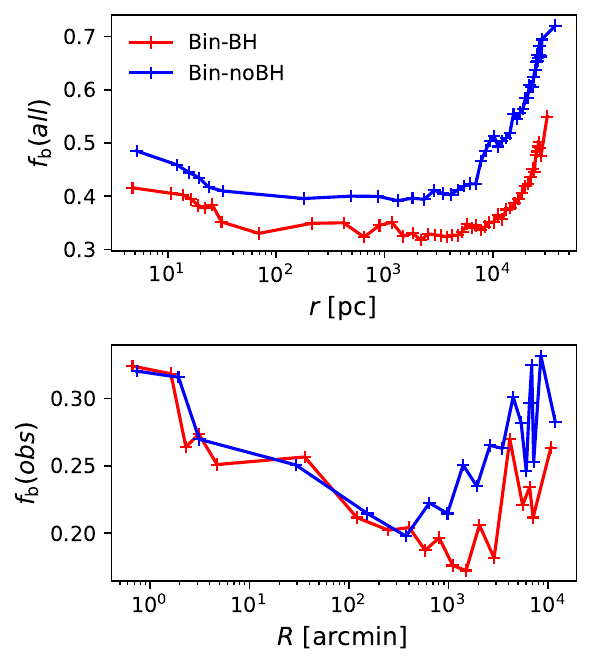}
    \caption{Upper panel: binary fractions of all objects along the 3D radial direction for the Bin-BH and Bin-noBH models; Lower panel: prediction for the observed binary fractions with an I-band magnitude range of 20.5 and 23 mag (corresponding to main sequence stars) and mass ratio $>0.6$.}
    \label{fig:rdist}
\end{figure}

Figure~\ref{fig:rdist} compares the radial distribution of the binary fraction ($\fbin$) for the Bin-BH and Bin-noBH models at 11.5~Gyr.

In the upper panel, the real $\fbin$ is plotted as a function of the 3D radial distance from the cluster center. 
Both models exhibit a similar trend, with a systematic offset of $\fbin$ along $r$. 
The central region of the cluster shows a higher $\fbin$ compared to the outer halo.
At the distant tail of the cluster, $\fbin$ experiences a significant increase.
This can be attributed to binaries that escaped from the cluster during the early stages of evolution, as they suffer fewer dynamical perturbations and have a higher chance of survival. 

The lower panel of Figure~\ref{fig:rdist} presents the predicted observed binary fraction as a function of projected distance. 
To identify binaries from the color-magnitude diagram, we assume that unresolved binaries with B-band magnitudes between 20.5 and 23 mag and a mass ratio above 0.6 can be detected. 
The B-band magnitudes for stars are generated by using \textsc{galevnb}.
Notably, $\fbin$ (obs) for both models is nearly identical within a projected distance up to 30 arcmin, unlike the real $\fbin$ for all binaries. 
The observed binary fraction $\fbin$ (obs) falls in the range of 0.2 to 0.3. 

\subsubsection{Half-year evolution of line-of-sight velocities}
\label{sec:dvr}

\begin{figure}
    \includegraphics[width=\columnwidth]{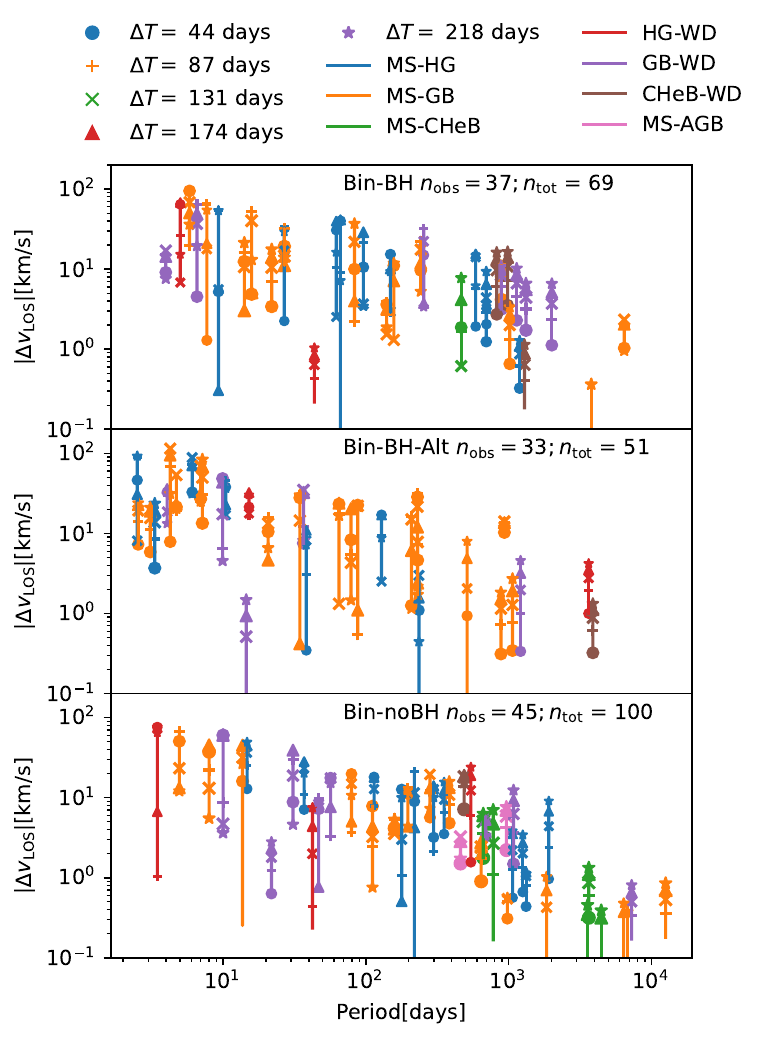}
    \caption{The line-of-sight velocity difference of binaries ($\dvr$) as a function of period for multi epochs of observation. The initial snapshots of the three models are chosen at $T=\Tmat$. Each binary type, classified according to the \textsc{sse} (Single Stellar Evolution) code, is represented by a different color. The stellar types include: MS (Main Sequence), HG (Hertzsprung Gap), GB (First Giant Branch), CHeB (Core Helium Burning), AGB (Asymptotic Giant Branch), and WD (White Dwarf).}
    \label{fig:pvdr}
\end{figure}
%

%

\begin{figure}
   \includegraphics[width=\columnwidth]{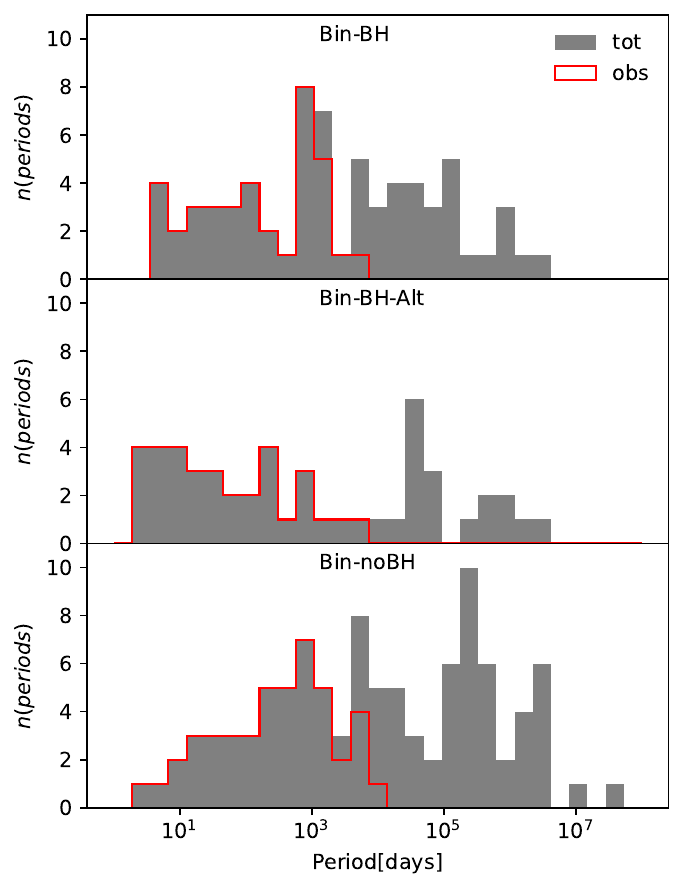}
   \caption{The number counts of bright binaries with post-main-sequence component for three models at . The legend "tot" include all binaries and the "obs" include only detectable binaries with $\dvr>0.3$~km/s.}
   \label{fig:pdvrhist}
\end{figure}

\begin{figure}
   \includegraphics[width=\columnwidth]{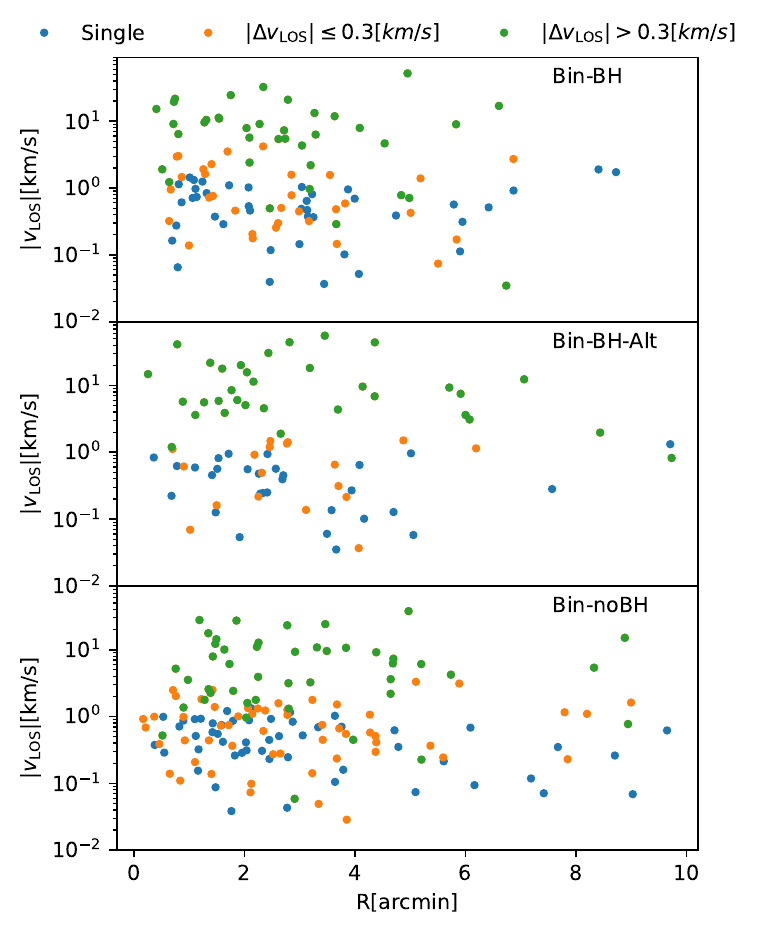}
   \caption{The line-of-sight velocities of individual bright stars and binaries are plotted, and detectable binaries with $\dvr>0.3$\,km/s are indicated as green dots.}
   \label{fig:sr}
\end{figure}

With high-resolution multi-epoch spectroscopic observations, it is possible to identify binaries by comparing the line-of-sight velocity changes ($\dvr$) over a span of approximately six months. 

The line-of-sight velocity $\vlos$ of an unresolved binary is the combination of two $\vlos$ of two components and is dominated by the brighter component.
Thus, the $\dvr$ values exhibit considerable variation during the multiple epochs of observation. 
These variations are determined by the periods, eccentricities, inclinations, and orbital phases of the binaries. 
Notably, larger variations are observed for short-period binaries, which could potentially aid in distinguishing these binaries from other effects that cause changes in  velocity. 
The baseline of approximately half a year is sensitive to a maximum period of $\sim10^4\,$ days.

We estimate  $\vlos$ of binaries by taking the I-band flux-weighted average of the $\vlos$ of the two components.
In Figure \ref{fig:pvdr}, we present the $\dvr$ versus period plot for observable unresolved binaries with $\dvr > 0.3~\text{km/s}$ and $R < 10$ arcmin after multiple epochs, respectively. 
We specifically select binaries with at least one bright (post-main-sequence) star component, and some binaries include white dwarfs. 
These bright stars have a luminosity in the HST $F555W$ filter brighter than 20 mag.
The three models (Bin-noBH, Bin-BH, and Bin-BH-Alt) exhibit observable binaries across a wide range of period distributions, spanning from 1 to $10^4$ days. 
The snapshots at $\Tmat$ (see the bottom panel of Figure~\ref{fig:ndp}) are chosen as the first epoch of observation. 
The choices of time intervals between epochs were chosen to be roughly equal space in half a year time interval, and the exact values are defined by the time step algorithm of the \textsc{petar} code.

The number of detectable binaries is similar for all three models, with the Bin-noBH model exhibiting slightly more binaries with periods above 3000 days.
This trend aligns with the period distributions shown in Figure \ref{fig:bp}, although some stochastic scatter may be present.

To assess the completeness of detectable binaries via multi-epoch observations of $\dvr$, we compare the number counts of detectable binaries and all bright binaries as a function of periods, as shown in Figure~\ref{fig:pdvrhist}.
For all models, periods up to $10^4$~days are detectable and all binaries with periods below $10^3$~days can be detected with multiple epochs. 
From Figure~\ref{fig:pvdr}, one binary in the Bin-BH model with a period between $10^3-10^4$~days has only one epoch that shows $\dvr>0.3$~km/s. 
A few binaries above $10^3$~days in the Bin-noBH models have epochs where $\dvr<0.3$~km/s, indicating that they might be missed if the observational epochs are limited to two.

The observed $\vlos$ of unresolved binaries does not represent the $\vlos$ of the center-of-mass of the binaries, which complicates the determination of the physically useful line-of-sight velocity dispersion ($\slos$).
A complete sample of detectable bright binaries with periods below $10^4$~days can mitigate this effect and significantly improve the determination of ($\slos$).
When binaries are detectable from multi-epoch observations, we can exclude them from the computation of $\slos$.
In our $N$-body model, we simulate the impact of excluding binaries with $\dvr>0.3$~km/s on the determination of $\slos$.

Figure~\ref{fig:sr} displays the individual line-of-sight velocities of bright stars ($\vlos$), undetectable bright binaries with $\dvr\leq 0.3$~km/s ($\slosb$), and detectable binaries with $\dvr > 0.3$\,km/s, aligned with the projected distance.
Most binaries with $\vlos>1$\,km/s are detectable, and thus, we can remove them for the calculation of $\slos$.

\begin{table}
    \centering
    \caption{The table displays the line-of-sight velocity dispersion ($\slos$) estimated from bright stars and binaries. The last column, $\slosvirial$, represents the estimation of $\slos$ based on Equation~\ref{eq:sigma_virial}, which serves as the unit for the other four columns. In particular, the column $\slossrc$ presents the $\slos$ value derived from single stars within $R<3$~arcmin (17~pc, approximately the $\rhn$). The remaining three columns depict $\slos$ within $R<10$ arcmin (58~pc), where $\sloss$, $\slosb$, and $\sloscb$ represent the $\slos$ values from only single stars, both single stars and binaries, and both single stars and undetectable binaries with $\dvr\leq 0.3$ km/s, respectively.
    }
    \begin{tabular}{cccccc}
    \hline
     Model & $\slossrc$ & $\sloss$ & $\slosb$ & $\sloscb$ & $\slosvirial$ \\
           & [$\slosvirial$] & [$\slosvirial$] & [$\slosvirial$] & [$\slosvirial$] & [km/s] \\
    \hline
     Bin-BH & 1.04 & 1.13 & 12.9 & 1.83 & 0.645\\
     Bin-BH-Alt & 1.01 & 1.05 & 22.9 & 1.33 & 0.528\\
     Bin-noBH & 1.02 & 0.815 & 8.81 & 1.27 & 0.729\\
    \hline
    \end{tabular}
    \label{tab:sigma}
\end{table}

Table~\ref{tab:sigma} demonstrates how removing detectable binaries improves the determination of $\slos$.
To have a consistent comparison among the three models, we scale the value of $\slos$ by the estimated 1-dimensional velocity dispersion $\slosvirial$ within $\rh$, assuming a virial equilibrium state of the cluster:
\begin{equation}
    \slosvirial \simeq \sqrt{\frac{G M}{6 \rh}}
    \label{eq:sigma_virial}.
\end{equation}
This normalization allows us to account for any differences in the overall dynamical state of the clusters and facilitates a more meaningful comparison of the $\slos$.

The presence of BHs affects the $\slos$ in the cluster center. 
To illustrate the difference between models with and without BHs, we calculate the $\slos$ of single stars within a projected distance of $R<3$~arcmin ($\slossrc$), which corresponds to the $\rhn$ (17~pc).
All three models exhibit similar values of $\slossrc$. Additionally, the $\slos$ values of single stars within a projected distance of $R<10$~arcmin (58pc), which includes stars outside the effective radius of the cluster, are similar to $\slossrc$, except the Bin-noBH model, which has a lower value.

Since the normalization factor $\slosvirial$ is different for the three models, and the observation cannot directly obtain $M$ and $\rh$, the difference in the observed estimates of $\slos$ for the three models may be larger than what we found in our simulations. 
This should be taken into consideration when interpreting the results and comparing them with observations.

The sample that includes all bright singles and binaries exhibits much larger dispersion values ($\slosb$) than the values ($\sloss$) of the sample containing only singles.
By excluding detectable binaries, the values ($\sloscb$) are significantly lower than $\slosb$, roughly 1.5-2 times of $\sloss$. 
This procedure helps to obtain more accurate estimates of $\slos$.


%

\subsubsection{Binaries with BHs}

The Bin-BH model  at 11.5 Gyr exhibits several binaries which contain one or two BHs (BwBHs), as depicted in Figure\ref{fig:denmap}.
It is important to investigate whether these BwBHs can be detected, serving as evidence for the existence of BHs.
Table~\ref{tab:BwBH} provides a summary of the parameters for these binaries, which include three types: BBHs, BH with MS (BH-MS), and BH with WD (BH-WD). 
Other types of BH-star binaries are not detected.

The presence of BBHs has also been illustrated in Figure~\ref{fig:abh}, with the possibility of some being detected by GW detectors.
Three BBHs are inside the clusters and the other three distribute in the tidal stream.

An interacting BwBH that contains an accreting BH primary and a non-BH secondary star is particularly interesting as a potential X-ray or radio source that could be detected, providing evidence for the presence of BHs in Pal~5.
Unfortunately, there is no BwBH that contains a bright post-main sequence star at 11.5~Gyr, only a few BH-MS and BH-WD exist.

We calculate the Roche lobe radius using Equation 53 from \cite{Eggleton1983,Hurley2002}, with the semi-major axis replaced by the peri-center distance $p$:
\begin{equation}
\frac{R_{\mathrm{RL2}}}{p} = \frac{0.49 q^{2/3}}{0.6 q^{2/3} + \ln{(1+q^{1/3}})}
\end{equation}
where $q=m_2/m_1$.
The original formula assumes a circular orbit, which misses the eccentric binaries where the accretion may occur at the peri-center separation. To account for this, we use the peri-center distance $p$ instead.
When the stellar radius of the secondary star ($R_2$) is greater than or equal to the Roche lobe radius ($R_{\mathrm{RL2}}$), the secondary star fills its Roche lobe, and the accretion process might result in observable radiation.

The $R_2/R_{\mathrm{RL2}}$ values of BH-MS binaries in our models are below $10^{-3}$, indicating that no accretion occurs in these cases.
The BH-WD binaries have the potential to become ultraluminous X-ray sources (ULXs).
Detailed studies of the dynamical formation scenarios for these ULXs in globular cluster environments have been conducted by \cite{Ivanova2010}.
One BH-WD binary in our simulations has a period of 2.5 days and a peri-center distance ($p$) of $2R_\odot$, located $\sim4.5\,$pc away from the cluster center.
The ratio $R_2/R_{\mathrm{RL2}}$ is approximately $\sim0.04$, which does not yet reach the criterion for accretion.

\begin{figure}
    \centering
    \includegraphics[width=\columnwidth]{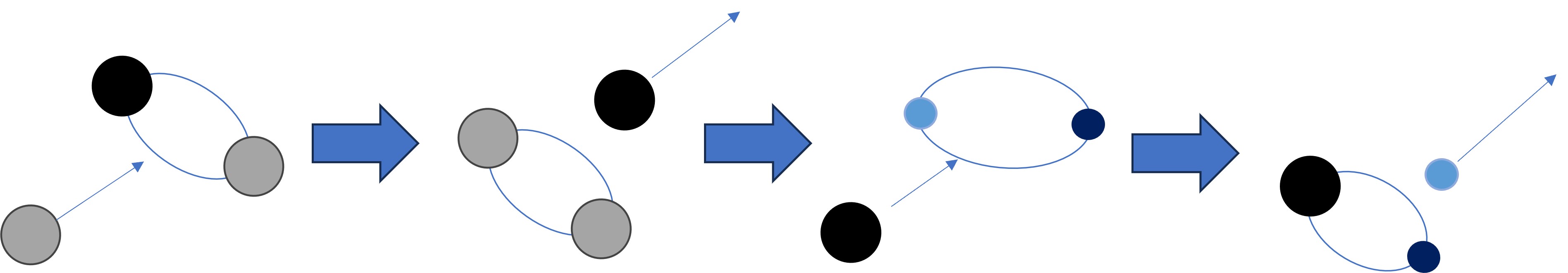}
    \caption{Illustration of the BH-MS formation process. The black and grey circles represent BHs, and the blue circles represent MS stars.}
    \label{fig:bhmsform}
\end{figure}

In our investigation of the BH-MS binaries, we have discovered that their formation occurs through a similar dynamical channel.
The MS star originates from a primordial binary of two MS stars (MS-MS).
The BH originates from a primordial binary of two massive stars, which forms a BBH.
The formation process of the BH-MS binaries in the Bin-BH model involves several steps:
\begin{enumerate}
    \item The BBH undergoes several interactions with other BHs in the cluster.
    \item After one of the BHs escapes from the cluster following a strong interaction with an intruder, it becomes a single BH.
    \item This single BH eventually encounters the MS-MS binary and participates in a binary exchange event.
    \item As a result of the binary exchange, the BH joins the MS-MS binary, forming the BH-MS binary.  
\end{enumerate}
The described process is visually illustrated in Figure~\ref{fig:bhmsform}.
The dynamical formation of BH-MS binaries in star clusters have been discussed in several works \citep{Kremer2018,DiCarlo2023,Rastello2023,Tanikawa2023}.

\begin{table*}
\caption{The parameters of BwBHs for the Bin-BH model at 11.5~Gyr. $m_1$ and $m_2$ denote the masses of the primary and secondary components, respectively; $p$ represents the peri-center distance; $R_2/R_{\mathrm{RL2}}$ indicates the secondary stellar radius relative to the Roche lobe overflow radius; and $r$ represents the distance of the binary from the cluster center.}
\label{tab:BwBH}
\begin{tabular}{cccccccc}
\hline
Type & $m_1[M_\odot]$ &  $m_2[M_\odot]$ &  period[days] &  $p[R_\odot]$ & eccentricity & $R_2/R_{\mathrm{RL2}}$ & $r$[pc] \\
\hline
BBH &39 & 27 & 5.9 & 41 & 0.26 & 8.1e-06 & 5.1e+03 \\
&37 & 30 & 5.9e+02 & 12 & 0.99 & 2.9e-05 & 9.1e+03 \\
&7.5 & 7.4 & 3.8 & 13 & 0.49 & 6.5e-06 & 9e+03 \\
&8.2 & 7.8 & 18 & 67 & 0.07 & 1.3e-06 & 5.4 \\
&7.6 & 7.6 & 24 & 61 & 0.29 & 1.4e-06 & 6.6 \\
&35 & 31 & 2.1e+04 & 6.1e+03 & 0.53 & 5.8e-08 & 8.7 \\
\hline
BH-MS & 21 & 0.66 & 1.8e+05 & 2.1e+04 & 0.45 & 0.00021 & 3.3 \\
& 16 & 0.71 & 3.3e+04 & 1.1e+03 & 0.90 & 0.0041 & 8.1 \\
& 13 & 0.68 & 4.8e+04 & 9e+03 & 0.33 & 0.00044 & 3.4 \\
& 15 & 0.21 & 1.1e+07 & 7.5e+04 & 0.86 & 2.7e-05 & 5.5 \\
\hline
BH-WD &8.4 & 1.1 & 1.4e+02 & 2.4e+02 & 0.01 & 0.00014 & 9.2 \\
&7.5 & 1 & 2e+02 & 2.7e+02 & 0.06 & 0.00012 & 13 \\
&16 & 0.74 & 1.8e+05 & 7.7e+03 & 0.78 & 8.7e-06 & 4.7 \\
&15 & 1 & 5.4e+06 & 9.6e+04 & 0.71 & 4.4e-07 & 6.4 \\
&8.2 & 0.52 & 2.5 & 2 & 0.87 & 0.039 & 4.5 \\
&15 & 0.69 & 3e+06 & 1e+05 & 0.52 & 6.7e-07 & 7.7 \\
\hline
\end{tabular}
\end{table*}

Although no observable events from interacting BwBH occur at 11.5\,Gyr, we can estimate the frequency of such events by collecting the interacting BwBHs recorded in the evolution of star clusters. 
The criterion to select interacting BwBHs are $R_2/R_{\mathrm{RL2}}\geq 1$. 
Events that occurred in the first 100~Myr are excluded, as they mostly involve primordial binaries that are not significantly affected by stellar dynamics. The results are summarized in Table~\ref{tab:BwBH-time}.

The Bin-BH and Bin-BH-Alt models have a dozen of such interacting BwBHs, including both primordial and dynamically formed BwBHs. 
The dynamically formed BwBHs contribute to approximately half of the interacting BwBHs.
The secondary stars involved in these BwBHs include several types, with one being BH-NS, which can trigger a GW merger.

The Bin-noBH model also includes 5 events, all of which consist of primordial binaries. 
Among these events, four are BH-MS binaries, and one is a BH-NS binary. 
Despite the high supernovae kick velocities in the Bin-noBH model, these binaries were strongly bound before the supernovae, and the random natal kick did not disrupt the binaries. 
Instead, the binaries escaped from the cluster after the kick.

In general, the formation rate of an interacting BwBH is estimated to be about one per 2~Gyr. 
Therefore, the possibility of detecting an interacting BwBH in the present-day Pal5 is practically zero.

The noBin-BH and Bin-noBH-F models do not exhibit any interacting BwBH events, and thus, they are not included in the table. 
One common feature of these two models is the absence of massive primordial binaries, which is different from all other models that have OB binary properties from \cite{Sana2012}. 
As a result, the possibility of dynamical formation of BwBHs is also low in these models. 
One important channel for the formation of interacting BwBHs is through the dynamical exchange of binary components after a close encounter between a BH and a binary. 
The lack of primordial binaries in these models suppresses this formation channel.

Multi-epoch observations of $\dvr$ can also be used to detect non-interacting BwBHs. 
For instance, utilizing multi-epoch MUSE spectroscopy, \cite{Giesers2018,Giesers2019} discovered three BwBHs in NGC3201. 
The stellar companions in these BwBHs have mass values of $0.6 - 0.8~M_\odot$. 
The four BH-MS binaries in the Bin-BH model at 11.5~Gyr have comparable companion masses. 
Therefore, it is possible to detect BHs in Pal~5 via multi-epoch observations of $\dvr$. 
However, due to the long periods of these binaries, a long-term observation plan (several years) is needed to accurately constrain the masses of the BHs. 
Despite the fact that these binaries are not $v_{\rm LOS}$ variable over a short baseline of a few months, they may still be found: they should appear as member stars according to their position in the CMD, parallax and propor motion, but they have a large $v_{\rm LOS}$ offset. 
A solar-type star orbiting a $15\,{\rm M}_\odot$ BH with a  $10^4\,$d period has an orbital velocity of $\sim25$\,km/s. This predicted signal is worth looking for.


\begin{table*}
\caption{The accretion events of BwBHs after 100~Myr. The "Primordial" column indicates whether the binary is primordial (formed during the initial star cluster formation) or dynamically formed (formed through interactions within the star cluster after its formation). The "Type" column indicates the combination of binary companions. The secondary stellar types involved in the accretion events include: MS, HG , GB, CHeB, AGB, HeHG (Hertzsprung Gap Naked Helium star), WD and NS (Neutron star).}
\label{tab:BwBH-time}
\begin{tabular}{rllrrrrrr}
\hline
\multicolumn{9}{c}{Bin-BH}\\
\hline
    Time[Myr] & Primordial  & Type & $m_1[M_\odot]$ & $m_2[M_\odot]$ & period[days] & $p[R_\odot]$ & eccentricity & $R_2/R_{\mathrm{RL2}}$ \\
\hline
109 & True & BH-HeHG & 6.7 & 0.92 & 1.1e+02 & 1.9e+02 & 2.560109e-05 & 1.0 \\
188 & True & BH-MS & 7.5 & 3.3 & 0.84 & 8.3 & 1.692295e-05 & 1.0 \\
268 & True & BH-AGB & 20 & 2.3 & 2.7e+03 & 2.3e+03 & 3.729159e-09 & 1.0 \\
861 & True & BH-WD & 6.3 & 0.0083 & 0.061 & 1.2 & 0.04394221 & 1.0 \\
5997 & False & BH-MS & 32 & 0.42 & 7e+05 & 0.38 & 0.9999964 & 9.0 \\
7141 & False & BH-MS & 18 & 0.2 & 9.9e+03 & 0.14 & 0.9999717 & 14.1 \\
7474 & True & BH-NS & 7.5 & 1.2 & 1.7e-08 & 5.8e-05 & 4.307228e-09 & 1.0 \\
\hline
\multicolumn{9}{c}{Bin-BH-Alt}\\
\hline
132 & True & BH-HeHG & 11 & 0.84 & 2e+02 & 3.3e+02 & 1.158092e-05 & 1.0 \\
134 & True & BH-HG & 6.8 & 4.1 & 3.6 & 14 & 0.3512435 & 1.5 \\
138 & True & BH-AGB & 20 & 1.6 & 8.6e+03 & 4.4e+03 & 0.09723035 & 1.1 \\
190 & True & BH-HG & 8.3 & 3.5 & 40 & 1.1e+02 & 0 & 1.0 \\
4125 & False & BH-MS & 17 & 0.34 & 1.1e+07 & 0.22 & 0.9999996 & 11.2 \\
\hline
\multicolumn{9}{c}{Bin-noBH}\\
\hline
116 & True & BH-MS & 9.2 & 2.9 & 1.1 & 10 & 6.648911e-05 & 1.0 \\
147 & True & BH-MS & 10 & 2.8 & 1.1 & 10 & 6.142399e-05 & 1.0 \\
159 & True & BH-MS & 8.2 & 2.7 & 1.1 & 9.7 & 0.0001996311 & 1.0 \\
209 & True & BH-NS & 7.5 & 1.5 & 1.7e-08 & 5.9e-05 & 2.710078e-08 & 1.0 \\
1232 & True & BH-MS & 2.5 & 0.99 & 0.48 & 3.9 & 3.125196e-05 & 1.1 \\
\hline
\end{tabular}
\end{table*}

\subsection{Color-magnitude diagram}
\label{sec:cmd}

\begin{figure*}
    \includegraphics[width=\textwidth]{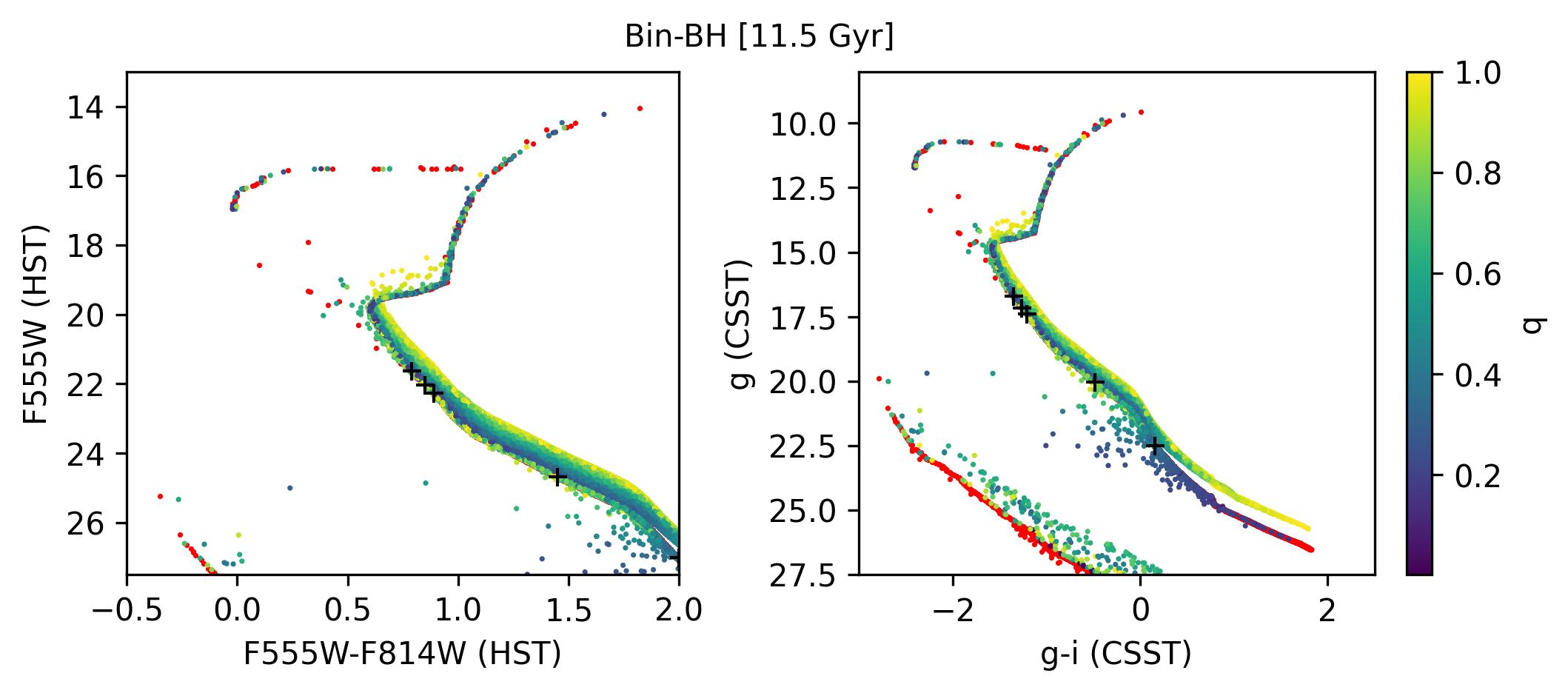}
    \caption{The color-magnitude diagram of the Bin-BH model at 11.5~Gyr. Red points are single stars. Other points are unresolved binaries where colors represent mass ratio ($q$). The black crosses are BH-MS binaries shown in Table~\ref{tab:BwBH}. The left panel corresponds to the HST F555W-F814W and F555W filters, while the right two panels correspond to the CSST g-i and g, and u-y and u filters, respectively.}
    \label{fig:cmd}
\end{figure*}

By utilizing the \textsc{galevnb} code, we can convert our simulation data into mock photometry. 
As an example, we present the color-magnitude diagram (CMD) of the Bin-BH model at 11.5~Gyr, using HST $F555W$ and $F814W$ filters, and CSST $g$ and $u-i$ filters (Figure\ref{fig:cmd}).

In the CSST filters, we observe binary stars distributed between the MS and WD sequence. 
These binaries consist of a WD and a low-mass main sequence star (LMS). 
Similar features in the CMD have been seen in $N$-body simulations by \citet{pang2022} (see figure 5 in \citealt{pang2022}) \footnote{In Pang et al. (2022), the CMD contained some horizontal strips of WD-LMS binaries, which was caused by a bug in the \textsc{petar} code. In that version of the code, some WDs had not evolved to the age of the snapshot, leading to this issue. However, in the CMD generated for this work, we have fixed this bug (in the commit on Jul 25, 2023 of the master branch of the \textsc{petar} code on GitHub), resulting in a more accurate representation of the stellar populations.}. 
In these binary systems, the luminosity is mainly dominated by the WD, as both components have very similar masses. 
They are considered as candidates for cataclysmic variable (CV) stars.

The CSST $g$-band magnitudes of WD and CV are below $26$ mag, while the corresponding HST F555W magnitudes are above $26$ mag. 
Therefore, CSST has the advantage of potentially detecting many WD and CV candidates in Pal~5.

We also highlight the BH-MS binaries shown in Table~\ref{tab:BwBH}. 
Among them, three have the HST F555W magnitude below 21 mag and the CSST $g$-band magnitude below 16 mag. 
If the multi-epoch spectroscopy observation can reach this magnitude limit, it is possible to detect these binaries via the observation of $\dvr$.

\subsection{Mass functions}
\label{sec:mf}

The present-day mass function of a star cluster is influenced by various factors, including the IMF, mass segregation, and tidal evaporation. 
To investigate the impact of primordial binaries and black holes (BHs) on the mass function, we compare the mass functions of our $N$-body models with the observed ones.
In order to make a meaningful comparison with the observed data, we select snapshots from our models that closely match the observed surface number density profile ($\Sigma(R)$), as shown in the lower panel of Figure~\ref{fig:ndp}.

It is important to consider the resolution limitations when comparing with observations. 
The widest binary in our models has a semi-major axis of approximately $1.8\times10^4$ AU. 
Given the distance to Pal~5, a spatial resolution of less than $1''$ is required to resolve this binary. 
The best resolution achievable by HST is around $0.05''$, which means that only a small fraction of wide binaries with periods above $1.4\times10^7$ days can potentially be resolved. 
Therefore, we assume that most binaries remain unresolved in observations and calculate their magnitudes by summing the fluxes of their two components.
Figure~\ref{fig:cmd} shows the color-magnitude diagram (CMD) of unresolved binaries, which appear redder and brighter compared to the single stars.


\begin{figure}
    \includegraphics[width=\columnwidth]{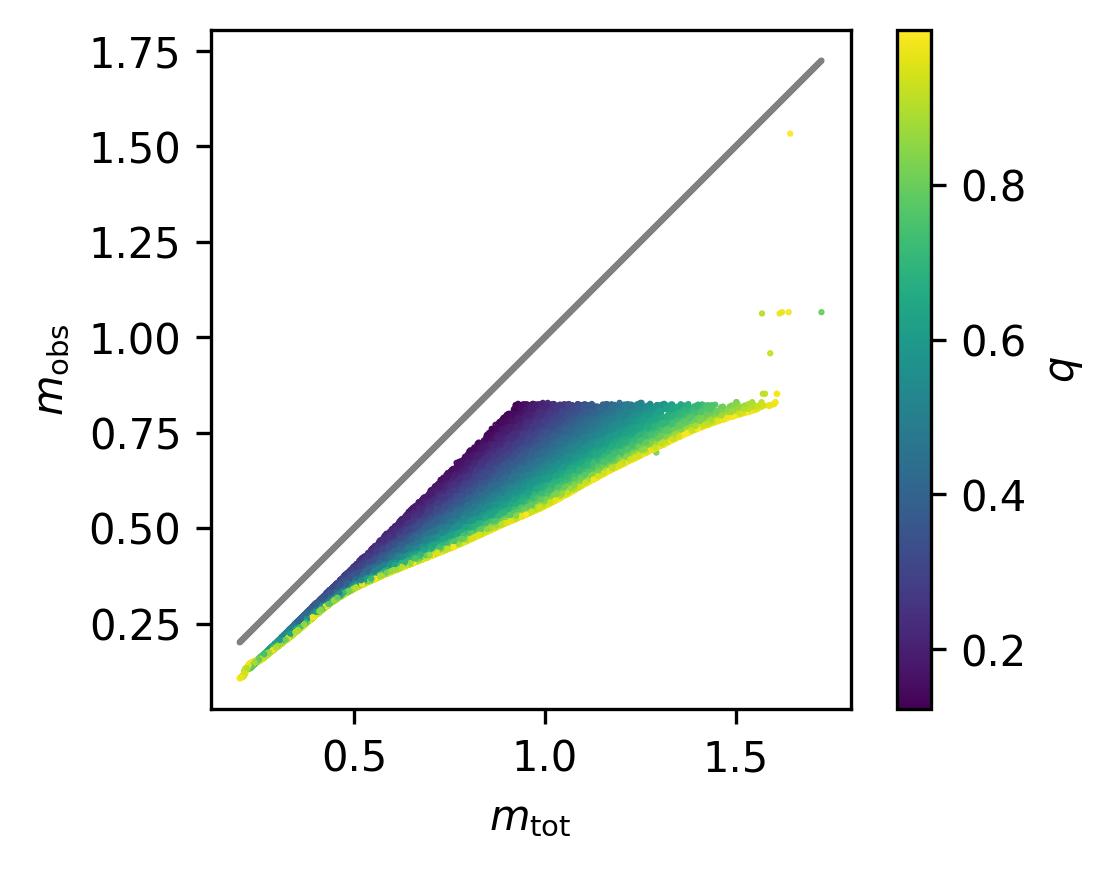}
    \caption{The total masses ($\mb$) v.s. the F555W-band flux-converted masses ($\mbv$) for main-sequence binaries of the Bin-BH model at 12~Gyr. The grey line shows the case of $\mb=\mbv$. Colors represent mass ratio ($q$).}
    \label{fig:mm}
\end{figure}

To investigate this effect, we compare the (actual) total masses ($\mb$) of binaries with the masses converted from their F555W-band magnitudes ($\mbv$).

For main-sequence binaries, we calculate the absolute F555W-band flux and then determine the mass of a single star that has the closest flux value, which serves as the converted mass $\mbv$. 
The comparison between $\mb$ and $\mbv$ is depicted in Figure~\ref{fig:mm}.

The difference between $\mb$ and $\mbv$ is highly sensitive to the mass ratio $q=m_1/m_2$ and luminosity ratio as well.
Here, the mass ratio $q$ is defined as the minimum mass divided by the maximum mass of the two components in a binary.
A higher $q$ leads to a larger difference between the $\mb$ and $\mbv$ 
 values. 
Consequently, the $\mbv$ of equal-mass unresolved binaries can be significantly lower than their true $\mb$. 

Furthermore, for binaries with the lowest $q$ values, there is a systematic offset between $\mb$ and $\mbv$. 
As a result, if unresolved main-sequence binaries cannot be distinguished from single stars, the total masses of all these binaries would be underestimated.

The offset between $\mb$ and $\mbv$ is determined by the minimum $q$.
There is a nonlinear relation between stellar luminosity ($L$) and mass ($m$). 
For MS stars in the mass range of 0.3-0.8~$M_\odot$, $L\propto m^4$, and thus, we can roughly estimate the relation between the total binary mass ($\mb$) and the binary mass used in the mass function estimation ($\mbv$) as follows:
\begin{equation}
\frac{\mbv}{\mb} \approx \frac{1+q^{4}}{1+q}.
\end{equation}
In our model, the minimum $q$ is about 0.12, which corresponds to a maximum $\mbv/\mb \approx 0.93$.

\begin{figure}
    \includegraphics[width=\columnwidth]{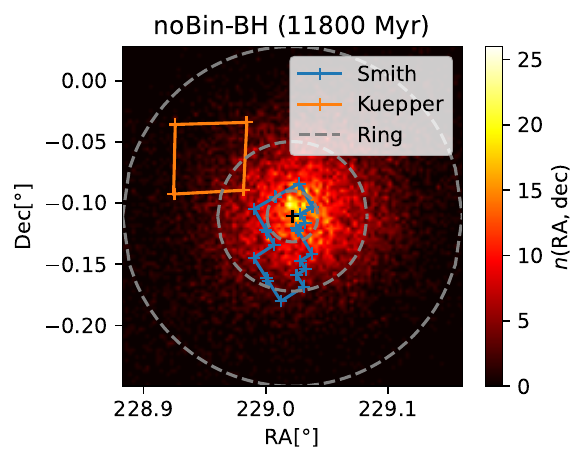}
    \caption{The 2-dimensional density map of the noBin-BH model at 11.8 Gyr. The color contours with solid lines represent the Smith and Kuepper fields, which have available HST data. The boundaries of the three ring radial bins are indicated by dashed grey circles. Two approaches are employed for selecting samples to measure the mass functions: 1) using the intersection between the Smith/Kuepper fields and the ring regions (referred to as "Field" regions); and 2) using only the ring regions themselves (referred to as "Ring" regions) to enhance statistical accuracy.}
    \label{fig:denmap}
\end{figure}

To compute the mass functions, we collect stars within the same observational fields used by the HST observation from the Smith field \citep{Grillmair2001}  and the Kuepper field \citep[unpublished; reported in][]{Baumgardt2023}, as shown in Figure~\ref{fig:denmap}. 
The center position of the star cluster model is defined as the centre-of-mass of stars located within the core of the star cluster. 
We adjust the center position to match the observed position of Pal~5.

The Smith field encompasses both the core and halo regions of Pal~5, while the Kuepper field covers the outer region. 
To investigate the radial dependence of the mass function in different regions of Pal~5, we divided the Smith and Kuepper fields into three radial bins. 
These bins correspond to different distances from the cluster center, allowing us to obtain mass functions as a function of radial distance.
The intersection between the two observational fields and the three radial bins (referred to "Field" regions) are used for selecting samples of stars.

It's important to note that due to the limited observational coverage and stochastic scatter, the comparison between the observed and modeled mass functions may be affected.
To improve statistical robustness, we also select stars for measuring the mass functions using only the three radial bins of the $N$-body models (referred to "Ring" regions).

By comparing the mass functions obtained from the $N$-body models and from the observed data, we can investigate the effects of primordial binaries and black holes on the mass function of Pal~5.

\begin{figure}
    \includegraphics[width=\columnwidth]{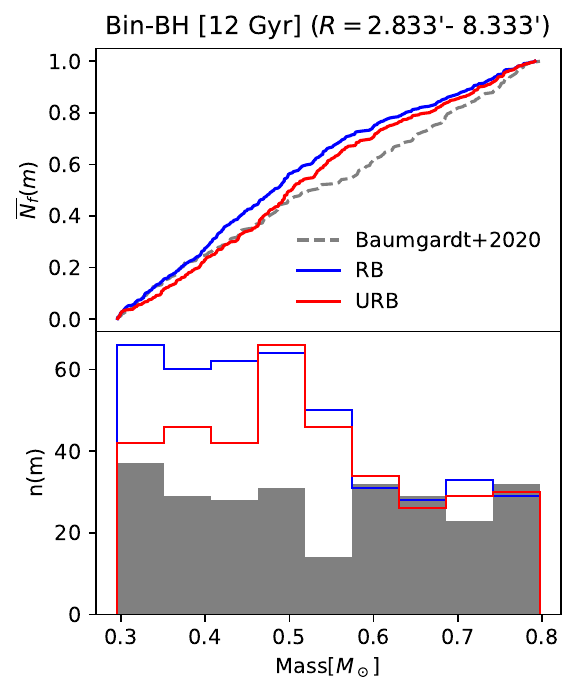}
    \caption{The mass functions of the Bin-BH model at 12~Gyr are presented in the Kuepper field, with the radial region indicated in the title. The observational data is shown as a reference. We compare different treatments of binaries in the mass function. "URB" indicates the use of $\mbv$ for mass estimation, and "RB" denotes the counting of masses for individual binary components. The upper panel displays the normalized cumulative counts, while the lower panel shows the normalized histograms.}
    \label{fig:mfres}
\end{figure}

We conducted an analysis to assess the impact of unresolved binaries on the determination of the mass function in the Kuepper field, using the Bin-BH model. 
The results are depicted in Figure~\ref{fig:mfres}. 
We considered two scenarios for the treatment of binaries in the mass function:
\begin{itemize}
\item RB (Resolved Binaries): All binaries are resolved, meaning that individual masses of binary components are counted in the mass function.
\item URB (Unresolved Binaries): $\mbv$ is utilized for mass estimation. This scenario represents a real observation where binaries are unresolved. 
\end{itemize}
The mass functions obtained from the RB and URB scenarios display steeper slopes compared to the observational mass function. 


\begin{figure*}
    \includegraphics[width=\textwidth]{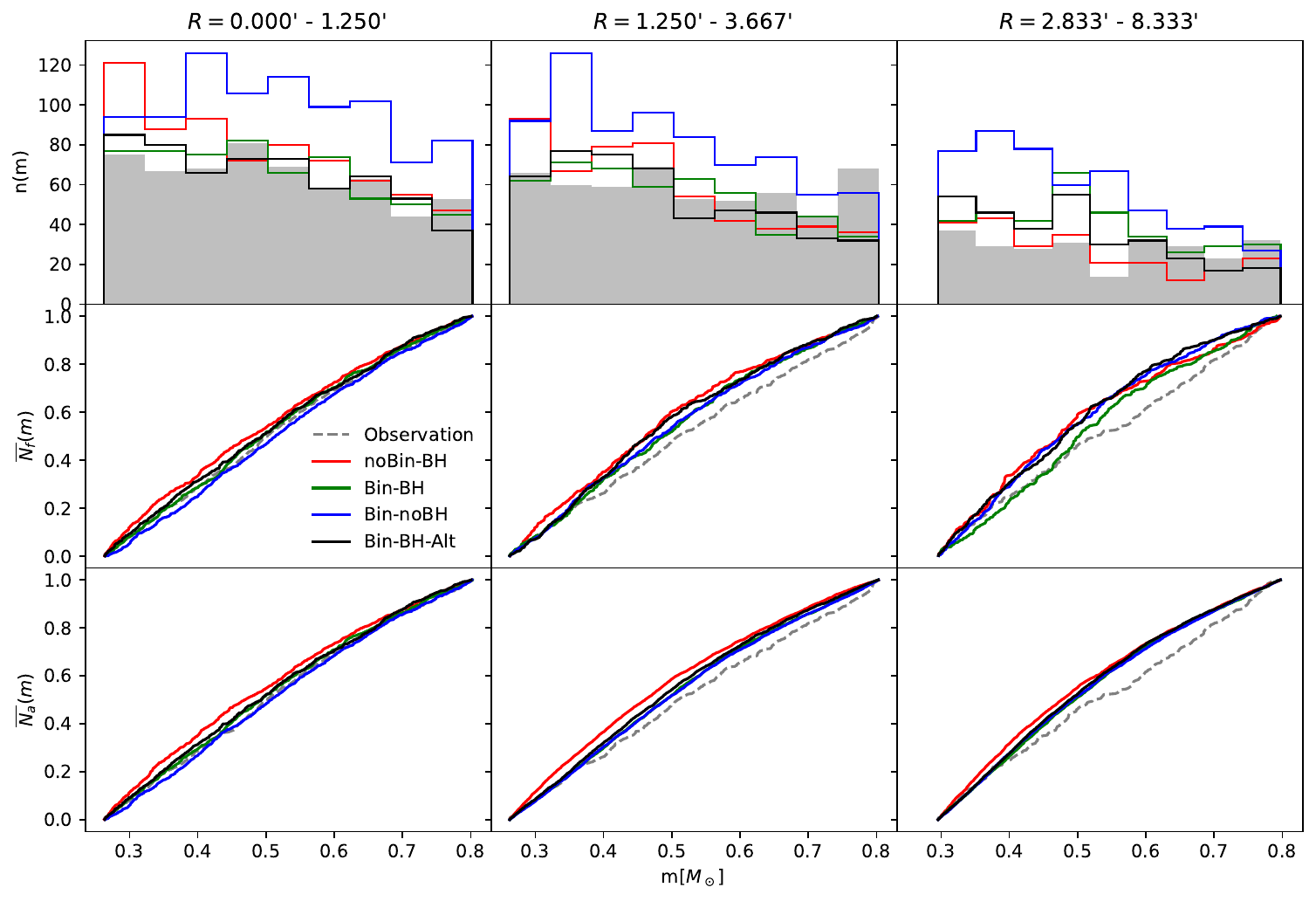}
    \caption{
    The mass functions of four $N$-body models in three radial bins, with the observational data shown as a reference. 
    The upper panel displays the number counts $n(m)$, the middle panel shows the normalized cumulative distribution $\cnm$ for the Field regions, and the lower panel shows the normalized cumulative distribution $\cnmb$ for the Ring regions.}
    \label{fig:mfall}
\end{figure*}

In Figure~\ref{fig:mfall}, we present a comparison between the mass functions obtained from the $N$-body models using the URB method and the observational data. 
The upper panel of Figure~\ref{fig:mfall} shows the number counts $n(m)$.
The $N$-body models exhibit a comparable number of stars within the three Field regions when compared to the observed data. 
The Bin-noBH model shows a slightly higher number of stars, indicating that a longer evolution time of more than 12 Gyr might be necessary for a better match. 
However, this slight discrepancy does not impact our comparison with the observed normalized counts.

The median and lower panels of Figure~\ref{fig:mfall} display the normalized cumulative distributions, $\cnm$, for the Field regions and $\cnmb$ for the Ring regions.
In the inner radial bin, no significant difference is observed when comparing $\cnm$ and $\cnmb$.
However, for the median and outer radial bins, a noticeable stochastic scatter is present in $\cnm$.
This scatter is particularly evident in the $\cnm$ of the Bin-BH model in the outer radial bin.
These findings suggest that the observational data may also exhibit similar scatter, and it is important to consider this when comparing the $N$-body model with the observational data.

The standard way to characterize a mass function is by using a power-law form given by the equation:
\begin{equation}
    \label{eq:mf}
    n(m) = C m^{-\alpha},
\end{equation}
where $C$ is a normalisation constant and $\alpha$ is the power-law index used for fitting.
We employ the fitting method outlined in \cite{Khalaj2013} to determine the statistical error accurately.
The formula for fitting $\alpha$ is:
\begin{equation}
    \label{eq:fit}
    \alpha = 1 + n \left [ \sum_{i=1}^{n}{\ln \frac{m_i}{m_{\mathrm{min}}}} - n \frac{\ln X}{1-X^{\alpha-1}} \right ]^{-1},
\end{equation}
where $n$ represents the total number of stars, $m_i$ is the mass of an individual star, $m_{\mathrm{min}}$ is the minimum mass of stars, and $X$ is the ratio of the maximum to the minimum masses of stars.
Iterative calculations are necessary to solve this fitting equation.
The corresponding error can be described as:
\begin{equation}
    \label{eq:error}
    \sigma(\alpha) = \frac{1}{\sqrt{n}} \left ( (\alpha-1)^{-2} - \ln^2{X} \frac{X^{\alpha-1}}{(1-X^{\alpha-1})^2} \right)^{-1/2} 
\end{equation}

The power-law indices of the mass functions ($\alpha$) obtained from fitting are summarized in Table~\ref{tab:alpha}. 
In the inner radial bin, the $\alpha$ values for the three Bin models are in rough agreement with the observational data, while the noBin-BH model shows a significantly higher $\alpha$. 
This result remains consistent when comparing the mass functions within the Field and the Ring regions.

In the middle and outer radial bins, all of the $N$-body models exhibit higher $\alpha$ values compared to the observational data. 
This discrepancy is more pronounced when considering the normalized cumulative distribution in the Ring regions ($\cnmb$).
These differences suggest that the $N$-body models exhibit more pronounced mass segregation than what is indicated by the observational data, although we need to take into account the potential stochastic scatter inherent in the observational data.
The presence of BHs does not appear to have a clear impact on the mass functions. 
The models incorporating primordial binaries exhibit better agreement with the observed data, particularly in the inner radial bin.

\begin{table*}
\centering
\caption{Fitting result of the power-law indices ($\alpha$) of the mass functions in different radial bins. The column labeled "region" distinguishes between the Smith and Kuepper fields (referred to "Field") and the ring regions (referred to "Ring").}
\label{tab:alpha}
\begin{tabular}{ccccccc}
\hline
R[arcmin] & region & Observation& noBin-BH& Bin-BH& Bin-noBH& Bin-BH-Alt\\
\hline
0.000 - 1.250& Field & 0.390$\pm$0.131& 0.835$\pm$0.119& 0.490$\pm$0.129& 0.198$\pm$0.107& 0.600$\pm$0.129\\
& Ring & & 0.882$\pm$0.104& 0.545$\pm$0.114& 0.300$\pm$0.093& 0.607$\pm$0.117\\
1.250 - 3.667& Field & 0.188$\pm$0.138& 0.987$\pm$0.135& 0.602$\pm$0.141& 0.637$\pm$0.115& 0.759$\pm$0.141\\
& Ring & & 0.997$\pm$0.052& 0.526$\pm$0.053& 0.508$\pm$0.044& 0.678$\pm$0.056\\
2.833 - 8.333& Field & 0.280$\pm$0.221& 1.174$\pm$0.226& 0.525$\pm$0.187& 1.140$\pm$0.154& 1.222$\pm$0.198\\
& Ring & & 1.127$\pm$0.061& 0.819$\pm$0.052& 0.840$\pm$0.046& 0.920$\pm$0.062\\
\hline

\end{tabular}
\end{table*}

\section{Limitations and Future Directions}
\label{sec:discussion}

\subsection{Uncertainty of initial condition}

Due to the computational expense, we are unable to explore the entire parameter space of the initial condition of Pal~5, resulting in several aspects not being addressed in this study.
These include assumptions regarding the properties of primordial binaries, the evolution of the Galaxy, the uncertainty associated with stellar evolution, the gravitational wave kicks following mergers of binary black holes (BBHs), and the realistic formation environment of the cluster.

In our study, we have adopted two extreme assumptions for the primordial binaries (Kroupa and FlatLog) with a 100\% initial binary fraction. 
However, these assumptions may not accurately reflect the true properties of primordial binaries in Pal~5. 
Nonetheless, Fig,~\ref{fig:bp} suggests that the initial period distribution has no significant impact on the survival fraction of binaries as a function of period, as long as the cluster possesses a similar initial density profile and orbit in the Galaxy. 
Furthermore, the evolution of the binary fraction ($\nhb$) can be utilized to derive the period evolution for different assumptions regarding the initial binary populations. 
By using a 100\% initial binary fraction, we also explore the maximum potential dynamical impact of primordial binaries. 
The wide range of periods considered allows us to investigate the behavior of hard and soft binaries with and without black holes (BHs).

Our model assumes a static Galactic environment, which is consistent with the setup employed in G21 to facilitate proper comparison. 
Incorporating a realistic time-dependent Galactic potential, which may be important to understand the density profile of the stream \citep*{Pearson2017}, is challenging due to the limited observational constraints on Galactic evolution. 
It is plausible that Pal~5 was formed in a significantly different Galactic environment, potentially leading to variations in mass loss and density evolution compared to our models. 
However, we believe that the overall trend driven by the presence of BHs should be similar.
Thus, our results offer a general perspective on how the existence of BHs impacts the binary populations.

The retention of BHs in clusters after supernovae remains an open question based on stellar evolution models. 
Our models do not consider gravitational wave kicks following BBH mergers, which could lead to an overprediction of massive BBHs with masses exceeding 100~$M_\odot$. 
Although such BBHs can influence the timescale of cluster disruption as shown in Figure~
\ref{fig:rht} and \ref{fig:abh}, their impact on the period distribution of binaries is limited since the hard-soft boundary is not determined by a single specific BBH.

The initial conditions of the clusters assume spherically symmetric Plummer models, similar to previous N-body simulations of GCs. 
However, the initial complexity of GC formation, including irregular cluster structures prior to achieving virial equilibrium and the presence of gas, may affect the binary populations during the gas-embedded phase.

\subsection{Observation of binaries}

In Section~\ref{sec:dvr}, we conducted an analysis to assess the feasibility of detecting binaries by measuring the radial velocity difference ($\dvr$) through multiple half-year observations. 
The maximum time-interval reaches half a year.
The results indicate that approximately 40 binaries could be identified, covering a period distribution ranging from a few to $10^4$ days. 
The model without BHs tends to exhibit a higher fraction of long-period binaries. 
While this observation cannot directly constrain the existence of BHs, it can provide insights into the presence of wide (long-period) binaries. 
Such information may be valuable in constraining the initial period distribution by utilizing the $\nhb$ values depicted in Figure~\ref{fig:bp}.

To obtain a stronger constraint on the existence of BHs, it is crucial to obtain additional observations of binaries in the period range around $10^5$ days, which has proven to be challenging thus far. 
Furthermore, it is necessary to observe binaries in different regions of Pal~5, including the inner region and the distant tail. 
Given the uncertainties associated with the properties of primordial binaries, assuming an initial period distribution becomes essential for constraining the density evolution based on the observed period distribution of present-day binaries. 
Notably, wide binaries disrupted within the dense cluster can survive along the low-density tidal tail. 
Therefore, the difference in the fraction of wide binaries inside the cluster and in the distant tail can help constrain both the initial period distribution of binaries and the density evolution of clusters, ultimately shedding light on the existence of BHs.

Another approach to constrain the BH population is by detecting BH-star binaries. 
We find four BH-MS binaries with relatively high MS masses, as shown in Table~\ref{tab:BwBH} and \ref{tab:BwBH-time} and also illustrated in Figure~\ref{fig:ramap}. 
Figure~\ref{fig:cmd} suggests that the CSST has the potential to detect CVs, thereby providing additional constraints on binaries with WDs.

Multi-epoch spectroscopic observations for $\dvr$ offer another possibility to detect non-interacting BH-star binaries. By utilizing this data, we can obtain better constraints on $\slos$, providing an indirect constraint on the dynamical impact from BHs in the cluster center.

\section{Conclusions}
\label{sec:conclusion}

In this study, we performed $N$-body simulations of the Galactic halo globular cluster Pal~5 with and without the inclusion of BHs, while considering a significant fraction of primordial binaries. 
Our main objectives were to investigate the influence of binaries and BHs on the cluster's dynamical evolution and to understand how the presence of BHs affects the binary populations within Pal~5. 
Additionally, we aimed to determine whether the observations of binary populations could provide indirect evidence for the existence of BHs in Pal~5.

Our findings indicate that the presence of primordial binaries has a noticeable but not drastic effect on the cluster's dynamical evolution, consistent with previous work \cite{Wang2022}. 
In models with BHs, the existence of primordial binaries alters the half-mass relaxation time ($\trh$) and reduces the number of BBHs that contribute to binary heating.
However, the influence on mass loss and radial evolution is more complex. 
Models with primordial binaries (Bin-BH and Bin-BH-Alt) exhibit shorter initial $\trh$ compared to models without primordial binaries (noBin-BH model). 
After 1 Gyr, the situation reverses due to larger half-mass radius ($\rh$) and lower total BH mass ($\Mbh$) in the Bin models. 
This trend changes again after 8 Gyr when a massive BBH forms in Bin-BH-Alt, accelerating the cluster's dissolution (see Figure~\ref{fig:abh}). 
Thus, the tidal dissolution time does not exhibit a simple dependence on the presence of primordial binaries.

In models without BHs and a low initial density (Bin-noBH and Bin-noBH-F), the evolution is more sensitive to the presence of primordial binaries compared to the BH models. 
Achieving a similar cluster at 11.5 Gyr requires a higher initial density in these cases.

Conversely, the assumption of BH existence significantly affects the population of wide binaries. 
Over long-term evolution, hard binaries are less affected by dynamical disruption. 
The fraction of hard binaries remains independent of the initial period distribution (Figure~\ref{fig:bp}). 
The remaining fraction of wide binaries depends on the evolution of the hard-soft boundary. 
The period distribution of models with BHs peaks at a shorter period compared to models without BHs, consistent with the hard-soft boundary. 
However, we find that not all wide binaries outside the hard-soft boundary are immediately disrupted. 
Many wide binaries outside this boundary can persist in the cluster for a long time. 
This suggests that the observation of wide binaries may not readily constrain the actual hard-soft boundary and be used to determine the cluster's density evolution history.

We have found that multi-epoch spectroscopic observations can detect most binaries with bright stars and periods below $10^4$~days. By excluding these binaries, the measurement of $\slos$ of bright stars can be significantly improved, providing better indirect constraints on the BH population through dynamical analysis.

Additionally, we have identified 4 BH-MS binaries in the Bin-BH model at 11.5~Gyr, which could potentially be detected using the same method, offering an additional possibility to provide evidence for the existence of BHs.

We also investigated how binaries and BHs influence the present-day mass function of Pal~5. 
Our results suggest that models with primordial binaries have mass function more consistent with the observational data, while the impact of BHs on the mass function is weak.
All $N$-body models exhibit mass segregation features that are not observed in the outer region of Pal~5.
However, it is important to consider the potential impact of stochastic scatter, which may influence the conclusions drawn from the comparison.
This indicates the need for alternative initial mass functions or additional observations of mass functions, with improved statistical precision, to better understand the underlying reasons for this discrepancy.

\section*{Acknowledgements}

L.W. thanks the support from the one-hundred-talent project of Sun Yat-sen University, the Fundamental Research Funds for the Central Universities, Sun Yat-sen University  (22hytd09).
L.W. and C.L. thank the support from the National Natural Science Foundation of China (NSFC) through grant 12073090.
L.W., C.L., X.P. and B.T. thank the support from NSFC through grant 12233013.
M.G. acknowledges financial support from the grants PID2021-125485NB-C22, EUR2020-112157,  CEX2019-000918-M funded by MCIN/AEI/10.13039/501100011033 (State Agency for Research of the Spanish Ministry of Science and Innovation) and SGR-2021-01069 grant (AGAUR).

\section*{Data Availability}

The simulations underlying this article were performed on the personal computing server of the first author.
The data were generated by the software \textsc{petar}, which is available in GitHub, at https://github.com/lwang-astro/PeTar.
The stellar evolution code \textsc{bse} is included in \textsc{petar}.
The \textsc{galpy} code for Galactic potential is available in GitHub, at https://github.com/jobovy/galpy. 
The initial conditions of star cluster models are generated by the software \textsc{mcluster}, which is available in GitHub, at https://github.com/lwang-astro/mcluster.
The \textsc{galevnb} code for mock photometry is available in GitHub, at https://github.com/xiaoyingpang/GalevNB.
The simulation data will be shared via private communication with a reasonable request. 



\bibliographystyle{mnras}
\bibliography{ref} 




%
%


\bsp	
\label{lastpage}
\end{document}